\documentclass[11pt, a4paper, twoside]{article}
\pdfoutput=1
 \usepackage[font=small, format=hang, margin={1cm, 1cm}]{caption}
\usepackage[a4paper, left=3cm,right=2cm]{geometry}
\usepackage{amsfonts} 
\usepackage{amsmath} 
\usepackage{amssymb}
\usepackage{mathabx}
\usepackage{mathtools}
\usepackage{relsize}
\usepackage{setspace}   
\usepackage{color}  
\usepackage{dutchcal}
\usepackage{wasysym}
\usepackage{booktabs}
\usepackage[utf8,applemac]{inputenc} 
\usepackage{tensor}
\usepackage{cite}
\usepackage{tikz}
\usepackage{graphicx}% Include figure files
\usepackage{subfig}		% Interface ÔøΩtendue pour objets flottants
\usepackage{pdfpages}
\usepackage{caption}
\usepackage[normalem]{ulem}
\usepackage{slashed}

\usepackage{epstopdf}
\usepackage{comment}
\numberwithin{equation}{section}
\usepackage{dcolumn}% Align table columns on decimal point
\usepackage{bm}% bold math
\usepackage{hyperref}

\newcommand{\bea}{\begin{eqnarray}}
\newcommand{\eea}{\end{eqnarray}}
\newcommand{\be}{\begin{equation}}
\newcommand{\ee}{\end{equation}}

\newcommand{\beqs}{\begin{eqnarray}}
\newcommand{\eeqs}{\end{eqnarray}}

\newcommand{\mn}{{\mu\nu}}
\newcommand{\p}{\partial}

\newcommand{\eps}{\epsilon}

\newcommand{\ndelta}{\delta\hspace{-0.50em}\slash\hspace{-0.05em} }
\newcommand{\Order}[1]{\mathcal{O}(r^{-#1})}

\newcommand{\pref}{\frac{\sqrt{q}}{16\pi G}}
\newcommand{\prefwg}{\frac{1}{16\pi G}}

\hypersetup{
    colorlinks,%
    citecolor=blue,%
    filecolor=blue,%
    linkcolor=blue,%
   urlcolor=blue,
   linktoc=page
}

\title{Superrotations}

\begin{document}

%\maketitle
%\abstract{[...]}

\setcounter{tocdepth}{2}

\begin{titlepage}

\begin{flushright}\vspace{-3cm}
{\small
%{\tt arXiv:yymm.nnnn} \\
\today }\end{flushright}
\vspace{0.5cm}

\begin{center}
{ \LARGE{\bf{Superboost transitions, refraction memory and\\ \vspace{3pt} super-Lorentz charge algebra}}}
 \vspace{5mm}

\vspace{2mm} 
\centerline{\large{\bf{Geoffrey Comp\`{e}re\footnote{e-mail: gcompere@ulb.ac.be}, Adrien Fiorucci\footnote{e-mail: afiorucc@ulb.ac.be}, Romain Ruzziconi\footnote{e-mail: rruzzico@ulb.ac.be}}}}

\vspace{2mm}
\normalsize
\bigskip\medskip
\textit{Universit\'{e} Libre de Bruxelles and International Solvay Institutes\\
CP 231, B-1050 Brussels, Belgium\\
\vspace{2mm}
}
\vspace{25mm}
%\vfil
%\pacs{04.70.Dy}

\begin{abstract}
We derive a closed-form expression of the orbit of Minkowski spacetime under arbitrary Diff$(S^2)$ super-Lorentz transformations and supertranslations. Such vacua are labelled by the superboost, superrotation and supertranslation fields. Impulsive transitions among vacua are related to the refraction memory effect and the displacement memory effect. A phase space is defined whose asymptotic symmetry group consists of arbitrary Diff$(S^2)$ super-Lorentz transformations and supertranslations. It requires a renormalization of the symplectic structure. We show that our final surface charge expressions are consistent with the leading and subleading soft graviton theorems. We contrast the leading BMS triangle structure to the mixed overleading/subleading BMS square structure. 

\end{abstract}

%\pacs{04.65.+e,04.70.-s,11.30.-j,12.10.-g}

\end{center}

%%%%%%%%%%%%%%%%%%%%%%%%%%%%%%%%%%%%%%%%%%%%%%%%%%%%%%%%%%%%%%%%%%%%%%%%%%%%%%%%%%%%%%%%

\end{titlepage}

\newpage
\tableofcontents

\section{Introduction and outline of the results}

The BMS group consisting of the infinite-dimensional commuting supertranslation subgroup and the Lorentz subgroup has been established as the asymptotic symmetry group of asymptotically flat spacetimes with original (Bondi, van der Burg, Metzner and Sachs) boundary conditions \cite{Bondi:1962px,Sachs:1962wk}. Supertranslation symmetry leads to a Ward identity in perturbative quantum gravity which is equivalent to Weinberg' soft graviton theorem \cite{He:2014laa}. The corresponding Noether charge is the flux of supertranslation charge at null infinity, which can be interpretated as a localized notion of energy on the celestial sphere \cite{Strominger:2013jfa}. Also, supertranslations shift a canonical field defined at the future and past of future null infinity $\mathcal I^+_\pm$, the supertranslation field \cite{Strominger:2013jfa}, whose finite differences encode the displacement memory effect \cite{Strominger:2014pwa}. This leads to a triangle relationship between supertranslation symmetry, Weinberg' soft graviton theorem \cite{Weinberg:1965nx} and the displacement memory effect \cite{Zeldovich:1974aa,Christodoulou:1991cr,Blanchet:1987wq,Blanchet:1992br,0264-9381-9-6-018}, see \cite{Strominger:2017zoo,Compere:2018aar} for reviews. 

A second subleading triangle has been suggested involving at its corners the subleading soft graviton theorem \cite{Cachazo:2014fwa}, the spin memory effect\footnote{Note that it has not (yet) been proven to be a memory effect in the sense that the observable can be expressed solely from the initial and final state. For another related effect, see \cite{Nichols:2018qac}.} \cite{Pasterski:2015tva} and super-Lorentz  symmetry\footnote{We find convenient to denote the extensions of the Lorentz transformations as the super-Lorentz transformations. Any 2-vector on the sphere can be decomposed into a divergence-free part and a rotational-free part. A super-Lorentz transformation whose pull-back on the celestial sphere is divergence-free is a superrotation. This generalizes the rotations. A super-Lorentz transformation whose pull-back on the celestial sphere is rotational-free is a superboost. This generalizes the boosts.} \cite{Barnich:2009se,Campiglia:2014yka} (see also \cite{deBoer:2003vf}). However, the nature of the relationship is more subtle at several levels. First, two distinct extensions of the original BMS group have been proposed as asymptotic symmetry groups of Einstein gravity at null infinity:
\begin{enumerate}
%\item[(i)] The original BMS group $SO(3,1) \ltimes \mathcal S $   \cite{Bondi:1962px,Sachs:1962wk}
\item[(i)] The Barnich-Troessaert group $(\text{Vir} \times \text{Vir}) \ltimes \mathcal S $ \cite{Barnich:2009se,Barnich:2010eb,Barnich:2011mi}  (see also \cite{deBoer:2003vf});
\item[(ii)] The Campiglia-Laddha group $\text{Diff}(S^2) \ltimes \mathcal S $ \cite{Campiglia:2014yka,Campiglia:2015yka}.
\end{enumerate}
In each case, $\mathcal S$ is the abelian subgroup of commuting supertranslations. In the first case, the Lorentz transformations are extended to meromorphic and anti-meromorphic transformations (with poles on the sphere), which thereby requires by consistency of the algebra that supertranslations with poles should also be considered. In the second case, the Lorentz transformations are extended to  arbitrary (smooth) diffeomorphisms on the 2-sphere and supertranslations are unchanged. 

The definition of a set of boundary conditions invariant under an asymptotic symmetry group is consistent if and only if the following set of conditions are met: (a) the asymptotic symmetries should preserve the boundary conditions; (b) all charges associated with asymptotic symmetries should be finite; (c) all charges should be well-defined (integrable).  In the particular case of boundary conditions defined at null infinity, since null infinity is permeable to energy flux, the third condition (c) has to be relaxed and a prescription to define the surface charge from the infinitesimal canonical charge has to be given \cite{Wald:1999wa}.

The boundary conditions leading to the first extension (i) of the BMS group cannot obey the second condition (b) assuming the standard bulk definition of surface charges \cite{Regge:1974zd,Iyer:1994ys,Barnich:2001jy}. Indeed, the singular supertranslation surface charges of the Kerr black hole diverge \cite{Barnich:2011mi}\footnote{More dramatically, the surface charges are linearly divergent in $r$ at the location of the meromorphic poles, as can be deduced from our analysis, see \eqref{divch}.}. A consistent phase space therefore requires a renormalization of the symplectic structure, consistently with the ambiguity of adding boundary terms to the symplectic structure \cite{Iyer:1994ys}, which leads to additional contribution to the surface charges \cite{Compere:2008us}. The boundary conditions proposed in \cite{Campiglia:2015yka} for the second extension (ii) obey neither the first condition (a)\footnote{The surface charges are defined using the boundary condition $C_{AB}=\mathcal{o}(u^{-1})$, which is not preserved by the action of the symmetry group due to the inhomogenous transformation law of $C_{AB}$ with a term of order $u^1$.} nor (b)\footnote{The boundary condition $C_{AB} = \mathcal{o}(u^{-1})$ still leaves the symplectic flux linearly divergent in $r$ except around the boundaries $\mathcal I^+_\pm$.}. Instead, more general boundary conditions are required which  lead to a radial divergence of the standard surface charges. Again, renormalization is necessary for boundary conditions admitting the symmetry group (ii). The need for a renormalization procedure can be most simply understood from the fact that both proposed symmetry groups modify the metric at leading order and in that sense are overleading\footnote{The Virasoro symmetries (i) also change the boundary metric on the sphere by adding singular poles.}. 

The subleading soft graviton theorem implies the Ward identities of Virasoro super-Lorentz symmetries \cite{Kapec:2014opa}. However, the converse is not true. Non-meromorphic super-Lorentz transformations are required in order to derive all instances of the subleading soft graviton theorem \cite{Campiglia:2014yka}. Instead, the Ward identities of Diff$(S^2)$ symmetry (labelled by 2 arbitrary functions on the sphere) are equivalent to the subleading soft graviton theorem \cite{Campiglia:2014yka}\footnote{Note that the subleading part of the supertranslation charge also takes a form similar to the super-Lorentz charge and its Ward identity is implied by the subleading soft theorem \cite{Conde:2016rom}. However, it is not clear how an equivalence can be obtained given that a supertranslation is labelled by a single function on the sphere.}. The 2 arbitrary functions on the sphere of a super-Lorentz transformation are parametrically equivalent to the two polarizations and the soft momentum of the soft graviton. The classical limit of this Ward identity is the conservation of a localized notion of angular momentum on the celestial sphere which is encoded in the Bondi angular momentum aspect \cite{Hawking:2016sgy}, itself also entirely determined on-shell by 2 arbitrary functions on the sphere. For these reasons, the most relevant symmetry group for the subleading infrared structure of general relativity is the symmetry group (ii). 

Our main objective is to propose a definition of renormalized phase space for the symmetry group (ii) and derive some of its structure. One can think of this renormalized phase space as an extended phase space which contains the standard phase space of BMS \cite{Bondi:1962px,Sachs:1962wk} containing \textit{e.g.} binary black hole mergers \cite{Blanchet:1985sp} and additional ``cosmic events'' which are usually discarded. The  solutions particular to the extended phase space include Robinson-Trautman spacetimes \cite{2003esef.book.....S} and their impulsive limit \cite{Penrose:1972aa,Nutku:1992aa,Podolsky:1999zr,Griffiths:2002gm}. Impulsive gravitational waves can be understood as a cosmic string decay which separates two Minkowski vacua related by a super-Lorentz transformation \cite{Griffiths:2002hj,Griffiths:2002gm,Strominger:2016wns}. With respect to the standard BMS phase space, the super-Lorentz transformations are outer symmetries \cite{Compere:2016jwb} in the sense that they are not tangent to the phase space but they are still associated with finite charges (see also \cite{Flanagan:2015pxa}). The existence of outer symmetries is sufficient to imply the existence of Ward identities at tree-level.

After a review of Einstein gravity in BMS gauge in Section \ref{sec:Bondi}, we derive in Section \ref{sec:vac} the closed-form expression of the vacua that carry a non-linear action of the extended Diff$(S^2) \times \mathcal S$ BMS group. This construction generalizes to arbitrary Diff$(S^2)$ super-Lorentz transformations the one of \cite{Compere:2016jwb,Compere:2016hzt}. It allows to define the canonical fields at null infinity that transform under the extended BMS group and that label inequivalent vacua. Among the super-Lorentz transformations, a particular role is played by the superboosts. The associated canonical variable, the superboost field, determines the leading part of the news tensor at $\mathcal I^+_\pm$ as the trace-free part of the stress-tensor of an Euclidean Liouville theory. 

In Section \ref{sec:transitions}, we use the identification of the superboost field to reinterpret the Robinson-Trautman spacetimes and impulsive gravitational waves as superboost transitions. We identify the class of observers around null infinity which display the refraction memory or velocity kick effect \cite{Podolsky:2002sa,Podolsky:2010xh,Podolsky:2016mqg}, depending whether one considers respectively null or timelike geodesics. Any gravitational wave leads to a velocity kick of probe objects because energy is transmitted from the gravitational wave to the probes \cite{1957Natur.179.1072B} (see also \cite{Zhang:2017rno,Zhang:2017geq,Zhang:2018srn,Zhang:2018gzn} and references therein). In contrast, the velocity kick/refraction memory that we describe here is specific to the observers close to null infinity and can be described in terms of superboost field transitions. We will also describe a new non-linear displacement memory effect at null infinity distinct from \cite{Zeldovich:1974aa,Christodoulou:1991cr,Blanchet:1987wq,Blanchet:1992br,0264-9381-9-6-018} which occurs in joined supertranslation and superboost transitions. 

In Section \ref{sec:phasespace}, we define an extended phase space invariant under the action of the generalized BMS group but which does not allow superboost transitions. We show that our final surface charge prescription reproduces in the standard BMS phase space the fluxes required for the leading and subleading soft graviton theorems following \cite{He:2014laa,Campiglia:2014yka}. Notably, we obtain a new expression for the angular momentum in the standard phase space which differs from the expressions given in \cite{Flanagan:2015pxa}, \cite{Hawking:2016sgy} or identified as the integrable charge in \cite{Barnich:2011mi}. We finally show that the generalized BMS charges obey the algebra under the modified Dirac bracket introduced in \cite{Barnich:2011mi}.

\vspace{4pt}\noindent {\bf Note added: } In the final stages of preparation of this manuscript we received \cite{Distler:2018rwu} where surface charges associated with super-Lorentz transformations are proposed that are consistent with the leading and subleading soft theorems. Their expressions for the charges in the standard BMS phase space, and in particular their expression for angular momentum, agree with ours.

%\vspace{4pt}\noindent {\bf Note added in v3: } Two algebraic mistakes and several typos were corrected. Eqs. (2.4), (2.23), (5.6), (5.31), (5.37), (5.68), (5.69), (5.70) were modified and the main text was adapted accordingly.  These corrections are detailed in the Erratum \cite{Compere:2020aaa}. 

\vspace{4pt}\noindent {\bf Note added in v3: } Two algebraic mistakes and several typos were corrected. Eqs. \eqref{qu}, \eqref{dM}, \eqref{puqab}, \eqref{Hxifin}, \eqref{infinitesimal charges}, \eqref{eq:Algebra}, \eqref{eq:XixiEval} were modified and the main text was adapted accordingly.  These corrections are detailed in the Erratum \cite{Compere:2020aaa}.

\section{A review of Einstein gravity in Bondi gauge}
\label{sec:Bondi}

In this section we set up our notation and compare them with the literature. We will mainly follow the conventions of \cite{Barnich:2010eb}. We will consider a solution space obeying fall-off conditions that are larger than required to define a consistent phase space. We will impose the remaining boundary conditions only in Section \ref{sec:BC2}.  The solution space is large enough to accomodate either the double copy Virasoro asymptotic symmetry group \cite{deBoer:2003vf,Barnich:2009se} or the Diff$(S^2)$ asymptotic symmetry group \cite{Campiglia:2014yka}. For simplicity, we do not consider the coupling to matter, see \cite{Flanagan:2015pxa} for a partial generalization.

\subsection{Bondi coordinates and assumptions}
\label{sec:BC1}

We choose a set of Bondi coordinates $(u,r,x^A)$ where $u$ labels null outgoing geodesic congruences, $r$ is a parameter along these geodesics, and $x^A$ are 2 coordinates on the 2-sphere. The most general $4$-dimensional metric can be written in this gauge as
\begin{equation}
\text{d}s^2 = \frac{V}{r} e^{2\beta} \text{d}u^2 -2e^{2\beta} \text{d}u \text{d}r + g_{AB} (\text{d}x^A - U^A \text{d}u)(\text{d}x^B - U^B \text{d}u).
\label{gauge condition 1}
\end{equation} 
Bondi gauge is reached by imposing the determinant condition
\begin{equation}
\partial_r \left(\frac{\det(g_{AB})}{r^4}\right) = 0,
\label{gauge condition 2}
\end{equation}
which singles out $r$ as the luminosity distance.

Each metric coefficient is provided with suitable fall-off conditions. Here, we assume that there is a polynomial fall-off in $r$ at least at second order in the asymptotic expansion for all components. We take 
\begin{equation}
\begin{split}
\frac{V}{r} &= \mathring{V} + \frac{2M}{r} + \mathcal{O}(r^{-2}) ,\\
\beta &= \frac{\mathring{\beta}}{r^2} + \mathcal{O}(r^{-3}), \\
g_{AB} &= r^2 q_{AB} + r C_{AB} +D_{AB}+ \mathcal{O}(r^{-1}) ,\\
U^A &= \frac{\mathring{U}^A}{r^2} -\frac{2}{3} \frac{1}{r^3} \left[ N^A - \frac{1}{2} C^{AB} D^C C_{BC} \right] + \mathcal{O}(r^{-4}) 
\end{split}
\label{fall-off}
\end{equation} 
where all functions appearing in the expansions of $\frac{1}{r}$ depend upon $u$ and $x^A$. All 2-sphere indices in \eqref{fall-off} are raised and lowered with $q_{AB}$, and $D_A$ is the Levi-Civita connection associated to $q_{AB}$. The determinant condition \eqref{gauge condition 2} imposes in particular that $q_{AB} C^{AB}=0$. $C_{AB}$ is otherwise completely arbitrary, and its time derivative $N_{AB} = \partial_u C_{AB}$ is the Bondi news tensor which describes gravitational radiation.

Furthermore, we impose the boundary condition
\begin{equation}
\sqrt{q} = \sqrt{\bar q}. \label{qu}
\end{equation}
where $\delta \bar q=0$ and $\p_u \bar q = 0$ which implies as a consequence of Einstein's equations that $\partial_u q_{AB} = 0$. This prevents evolution among distinct boundary metrics. It is possible to relax this restriction but the expressions become lengthy and we will not derive them here. We refer the reader to Barnich-Troessaert \cite{Barnich:2010eb} for a partial generalization where all the dependence in $u$ of the boundary metric is in an overall conformal factor.

\subsection{Equations of motion}

Einstein's equations imply
\begin{equation}
\mathring{V} = - \frac{1}{2} \mathring{R}, \qquad
\mathring{\beta} = -\frac{1}{32} C^{AB}C_{AB}, \qquad
\mathring{U}^A = - \frac{1}{2} D_{B}C^{AB},\qquad D_{AB} =  \frac{1}{4}q_{AB}C^{CD}C_{CD} , \label{Ein2}
\end{equation} 
where $\mathring{R}$ is the Ricci scalar of $q_{AB}$. Einstein's equations are then fully obeyed at this order in the Bondi expansion except for the following two additional constraints: 
\begin{align}
\partial_u M &= - \frac{1}{8} N_{AB} N^{AB} + \frac{1}{4} D_A D_B N^{AB} + {\frac{1}{8} D_A D^A \mathring{R}},\label{duM} \\ 
\partial_u N_A &= D_A M + \frac{1}{16} D_A (N_{BC} C^{BC}) - \frac{1}{4} N^{BC} D_A C_{BC} \nonumber \\
&\quad -\frac{1}{4} D_B (C^{BC} N_{AC} - N^{BC} C_{AC}) - \frac{1}{4} D_B D^B D^C C_{AC} \label{EOM1} \\
&\quad + \frac{1}{4} D_B D_A D_C C^{BC} + { \frac{1}{4} C_{AB} D^B \mathring{R}}. \nonumber
\end{align}
Here $M(u,x^A)$ is the Bondi mass aspect, $N_A(u,x^B)$ is the angular momentum aspect. Concerning this quantity, our conventions are those of Barnich-Troessaert \cite{Barnich:2010eb,Barnich:2011mi} (also followed by \cite{Distler:2018rwu}), but differ from those of Flanagan-Nichols $(FN)$ \cite{Flanagan:2015pxa} and Hawking-Perry-Strominger $(HPS)$ \cite{Hawking:2016sgy}. Here is the dictionary to match the different conventions:
\begin{align}
N_A^{(FN)} &= N_A + \frac{1}{4} C_{AB} D_C C^{BC} + \frac{3}{32} \partial_A (C_{BC} C^{BC}) ,\label{chgtFN}\\
N_A^{(HPS)} &= N_A^{(FN)} - u D_A M.
\end{align}

\subsection{Residual diffeomorphisms}
The infinitesimal residual diffeomorphisms $\xi^\mu \partial_\mu$ preserving Bondi gauge and the constraint \eqref{qu} are given by 
\begin{equation}
\begin{split}
\xi^u &= f(u,x^A) ,\\
\xi^A &= Y^A(u,x^B) + I^A, \quad I^A = - D_B f \int_r^\infty \text{d}r' (e^{2\beta} g^{AB}), \\
\xi^r &= -\frac{1}{2} r (D_A Y^A + D_A I^A - U^A D_A f ),
\end{split} \label{eq:BMSvectors}
\end{equation}
with $\partial_r f = \partial_r Y^A = 0$. The additional fall-offs \eqref{fall-off} require that
\begin{equation}
\begin{split}
\mathcal{L}_\xi g_{uA} = \mathcal{O}(r^{0}) &\Rightarrow \partial_u Y^A = 0 \Leftrightarrow Y^A = Y^A (x^B),\\
\mathcal{L}_\xi g_{ur} = \mathcal{O}(r^{-2}) &\Rightarrow \partial_u f = \frac{1}{2}D_A Y^A \Leftrightarrow f = T(x^B) + \frac{u}{2}D_A Y^A ,
\end{split}
\label{fAndY}
\end{equation}
and nothing else. We can perform the radial integration in (\ref{eq:BMSvectors}) to get a perturbative expression of the infinitesimal residual diffeomorphisms using \eqref{Ein2}:
\begin{align}
\xi^u &= f ,\\
\xi^A &= Y^A - \frac{1}{r} D^A f + \frac{1}{r^2}  \left( \frac{1}{2} C^{AB} D_B f \right) + \frac{1}{r^3} \left( -\frac{1}{16} C_{BC}C^{BC} D^A f \right)  + \mathcal{O}(r^{-4}) ,\\
\xi^r &= -\frac{1}{2} r D_A Y^A + \frac{1}{2} D_A D^A f + \frac{1}{r} \left( -\frac{1}{2} D_A C^{AB} D_B f - \frac{1}{4} C^{AB} D_A D_B f \right) + \mathcal{O}(r^{-2}). \label{eq:XiR}
\end{align}
The residual diffeomorphisms are spanned by \textit{arbitrary} $\text{Diff}(S^2)$ super-Lorentz transformations generated by $Y^A (x^B)$ and by (smooth) supertranslations generated by $T(x^A)$. We will therefore denote them as $\xi(T,Y)$. 

\subsection{Commutator algebra}

In order to obtain the algebra of infinitesimal residual diffeomorphisms under the Lie bracket $[\cdot,\cdot]$, it is sufficient to consider the leading order vectors $\xi(T,Y) = f \partial_u + Y^A \partial_A$, where $f$ and $Y^A$ satisfy \eqref{fAndY}. Defining 
\begin{equation}
[\xi(T_1,Y_1),\xi(T_2,Y_2)] = \xi (T_{[1,2]},Y_{[1,2]})
\end{equation}
we find 
\begin{equation}
\begin{split}
T_{[1,2]} &= Y_1^A D_A T_2 + \frac{1}{2} T_1 D_A Y^A_2 - (1\leftrightarrow 2) , \\
Y_{[1,2]}^A &= Y^B_1 D_B Y_2^A - (1\leftrightarrow 2).
\end{split}
\end{equation}
This defines the \textit{generalized BMS algebra} $\mathfrak{g}$. It consists of the semi-direct sum of the diffeomorphism algebra on the celestial $2$-sphere $\mathfrak{diff}(S^2)$ and the abelian ideal $\mathfrak{s}$ of supertranslations, consisting of arbitrary smooth functions on the $2$-sphere.
\begin{equation}
\mathfrak{g} = \mathfrak{diff}(S^2) \oleft \mathfrak{s}.
\end{equation}
As in \cite{Barnich:2010eb}, it can be checked that, taking \eqref{fAndY} into account, the bulk vectors \eqref{eq:BMSvectors} form a faithful representation of that algebra for the modified Lie bracket
\begin{equation}
[\xi_1,\xi_2]_M = [\xi_1,\xi_2] - \left( \delta^g_{\xi_1} \xi_2 - \delta^g_{\xi_2} \xi_1 \right),
\end{equation}
where $\delta^g_{\xi_1} \xi_2$ denotes the variation of $\xi_2$ caused by the Lie dragging along $\xi_1$ of the metric contained in the definition of $\xi_2$. 

\subsection{Representation on the solution space}
The vectors \eqref{eq:BMSvectors} preserve the solution space in the sense that infinitesimally
\begin{equation}
\mathcal{L}_{\xi(T,Y)} g_{\mu\nu} [\phi^i] = g_{\mu\nu} [ \phi^i + \delta_{(T,Y)} \phi^i ] - g_{\mu\nu} [ \phi^i ] 
\end{equation}
where $\phi^i = \{ q_{AB}, C_{AB}, M, N_A \}$ denotes the collection of relevant fields that describe the metric in Bondi gauge. The action of the vectors preserve the form of the metric but modify the fields $\phi^i$, in such a way that the above equation is verified. We can show that
\begin{align}
\delta_{(T,Y)} q_{AB} &= 2 D_{(A} Y_{B)} - D_C Y^C q_{AB},\label{dqAB} \\
\delta_{(T,Y)} C_{AB} &= [f \partial_u + \mathcal{L}_Y - \frac{1}{2} D_C Y^C ] C_{AB} - 2 D_A D_B f + q_{AB} D_C D^C f,\label{dCAB}\\
\delta_{(T,Y)} N_{AB} &= [f\partial_u + \mathcal{L}_Y] N_{AB} - (D_A D_B D_C Y^C - \frac{1}{2} q_{AB} D_C D^C D_D Y^D),\label{dNAB}\\
\delta_{(T,Y)} M &= [f \partial_u + \mathcal{L}_Y + \frac{3}{2} D_C Y^C] M - \frac{1}{2} D_A f D^A \mathring{V} + \frac{1}{4} N^{AB} D_A D_B f \nonumber \\
&\quad + \frac{1}{2} D_A f D_B N^{AB}   + \frac{1}{8}D_A D_B D_C Y^C C^{AB}  , \label{dM}\\
\delta_{(T,Y)} N_A &= [f\partial_u + \mathcal{L}_Y + D_C Y^C] N_A + 3 M D_A f - \frac{3}{16} D_A f N_{BC} C^{BC} + \frac{1}{2} D_B f N^{BC} C_{AC} \nonumber \\
&\quad - \frac{1}{32} D_A D_B Y^B C_{CD}C^{CD} + \frac{1}{4} (D^B f \mathring{R} + D^B D_C D^C f) C_{AB} \nonumber \\
&\quad - \frac{3}{4} D_B f (D^B D^C C_{AC} - D_A D_C C^{BC}) + \frac{3}{8} D_A (D_C D_B f C^{BC}) \nonumber \\
&\quad + \frac{1}{2} (D_A D_B f - \frac{1}{2} D_C D^C f q_{AB}) D_C C^{BC}.   \label{dNA}
\end{align} 
Note that the boundary Ricci scalar $\mathring{R}$ (or $\mathring V$) transforms as
\begin{equation}
\delta_{(T,Y)} \mathring{R} = Y^A D_A \mathring{R} + D_A Y^A \mathring{R} + D^2  D_B Y^B.
\end{equation}

\section{Vacuum structure}
\label{sec:vac}

A special role is played by the action of the generalized BMS group on the Minkowski metric. The orbit of Minkowski spacetime under the BMS group is defined as the class of Riemann-flat metrics obtained by exponentiating a general BMS transformation starting from Minkowski spacetime as a seed. The subset of this orbit where only supertranslations act are the non-equivalent vacua of asymptotically flat spacetimes which are characterized, contrary to Minkowski spacetime, by non-vanishing super-Lorentz charges while all Poincar\'e charges remain zero \cite{Compere:2016jwb}. In this standard case, the exponentiation leads to a single fundamental field labeling inequivalent vacua: the supertranslation field $C(x^A)$. The displacement memory effect is a transition among vacua mediated by gravitational or other null radiation which effectively induces a supertranslation of $C$ \cite{Strominger:2014pwa}.

For the double copy Virasoro asymptotic symmetry group, this exponentiation leads to two fundamental fields: the supertranslation field and what we will call the superboost or Liouville field $\Phi$. The corresponding solution in Bondi and Newman-Unti gauges was constructed in \cite{Compere:2016jwb}. Here, we extend the construction to finite \text{Diff}$(S^2)$ super-Lorentz transformations following methods similar to the Appendix of \cite{Compere:2016hzt}. The corresponding boundary fields will also be the supertranslation $C$ and superboost $\Phi$ fields, complemented by an additional superrotation field $\Psi$. In order to understand the memory effects associated with super-Lorentz transformations, we therefore start by deriving the structure of the vacua.

\subsection{Generation of the vacua}
\label{Generation of the vacua}

We start from the Minkowski metric written in complex plane coordinates:
\begin{equation}
ds^2 = -2\text{d}u_C \text{d}r_C + 2 r_C^2 dz_C d\bar{z}_C. 
\end{equation}
We define the background structures
\begin{equation}
\gamma_{ab} = \left[ \begin{array}{cc}
0 & 1 \\ 
1 & 0
\end{array} \right], \quad \epsilon_{ab} = \left[ \begin{array}{cc}
0 & 1 \\ 
-1 & 0
\end{array} \right]
\end{equation}
with inverse $\gamma^{ab} = \gamma_{ab}$, $\epsilon^{ab} = \epsilon_{ab}$. The goal is to introduce a diffeomorphism to Bondi gauge $(u_C,r_C,z_C,\bar{z}_C)\to (u,r,z,\bar{z})$ that exponentiates $\text{Diff}(S^2)$ super-Lorentz transformations and supertranslations. Requiring that $(u,r,z,\bar z)$ are Bondi coordinates leads to 2 conditions:
\begin{itemize}
\item[$\rhd$] The coordinate $r$ parametrizes null radial geodesics: $g_{rr} = g_{rA} = 0$.
\item[$\rhd$] $r$ represents the luminosity distance in the sense of Sachs, so \\$\partial_r (r^{-4} \det g_{AB}) = 0$ (where $g_{AB} = g_{\mu\nu} \nabla_A x^\mu \nabla_B x^\nu$).
\end{itemize}
The first condition yields
\begin{align}
r_C &= r_C (r,u,z^c), \\
u_C &= W(u,z,\bar{z}) - r_C^{-1} \gamma_{ab} H^a (u,z^c) H^b (u,z^c), \\
z_C^a &= G^a (z^c) - r_C^{-1} H^a (u,z^c), \quad
H^a (u,z^c) = - D_G^{-1} \epsilon^{ab} \gamma_{bc}\epsilon^{AB} \partial_A W \partial_B G^c
\end{align}
where $D_G = \det (\partial_A G^b) = \frac{1}{2!}\epsilon_{ab}\epsilon^{AB}\partial_A G^a \partial_B G^b$. The second condition fixes the functional dependence of $r_C$ as
\begin{align}
r_C (r,u,z^c) &= R_0 (u,z^c) + \sqrt{\frac{r^2}{(\partial_u W)^2} + R_1 (u,z^c)}.\label{rc}
\end{align}
Here $R_0$ and $R_1$ are respectively obtained by requiring that the determinant condition is obeyed up to second and third order in $1/r$. The information at subleading orders is propagated with the power expansion of the functional dependence in $r$. 

Requiring that $g_{uu}$ is finite in $r$ and restricting the boundary metric as \eqref{qu}, we have to impose that $\partial^2_u W = 0$, so $W$ is at most linear in $u$. Moreover, regularity implies that it is nowhere vanishing. Therefore,
\begin{equation}
W(u,z^c) =  \exp \left[\frac{1}{2}\Phi (z,\bar z)\right] (u + C(z,\bar z)).
\label{Wform}
\end{equation}
%We now anticipate Section \ref{sec:BC2} and impose the  boundary condition $\sqrt{q}=\sqrt{\bar{q}}$ where $\bar{q}_{AB}$ is the unit metric on the sphere. For definiteness, we choose $z^A$ to be the stereographic coordinates of $S^2$, 
%\begin{equation}
%\bar{q}_{AB} dz^A dz^B = 2 \gamma_s dz d\bar{z}, \qquad \gamma_s = \frac{2}{(1+z\bar z)^2}. 
%\label{UnitRound}
%\end{equation} 
%The boundary condition is then equivalent to fixing the Jacobian of the coordinate transformation as 
%\bea
%D_G = \gamma_s e^{\Phi} . 
%\eea
Expanding $g_{AB}$ in powers of $r$ as in \eqref{fall-off}, we can read the boundary metric as 
\begin{equation}
q_{AB} =q_{AB}^{\text{vac}} \equiv e^{-\Phi} \partial_A G^a \partial_B G^b \gamma_{ab}.
\label{qABform}
\end{equation}
It is indeed the result of a large diffeomorphism and a Weyl transformation. The shear $C_{AB}$ is found to be the trace-free part (TF) of the following tensor
\begin{equation}
C_{AB} = C_{AB}^{\text{vac}} \equiv \left[ \frac{2}{(\partial_u W)^2} \partial_u \left( D_A W D_B W \right) - \frac{2}{\partial_u W} D_A D_B W \right]^{\text{TF}}.
\label{CABGeneral}
\end{equation}
Introducing \eqref{Wform}, it comes 
\begin{equation}
C^{\text{vac}}_{AB}[\Phi,C] =  (u +C) N^{\text{vac}}_{AB} + C^{(0)}_{AB}, \quad \left\lbrace \begin{array}{ccl}
%N^{\text{vac}}_{AB} & = & \left[ \frac{4}{W_1^2} D_A W_1 D_B W_1 - \frac{2}{W_1} D_A D_B W_1 \right]^{\text{TF}} \\ 
N^{\text{vac}}_{AB} & = & \left[\frac{1}{2}D_A \Phi D_B \Phi - D_A D_B \Phi  \right]^{\text{TF}} \label{PhiL} \\ 
C^{(0)}_{AB} & = &   - 2 D_A D_B C + q_{AB} D^2 C.
\end{array} \right.
\end{equation}
We find that all explicit reference on $\gamma_{ab}$ or $G^a$ disappeared. Moreover, the news tensor of the vacua $N^{\text{vac}}_{AB}$ is only built up with $\Phi$. It can be checked that the boundary Ricci scalar is given in terms of $\Phi$ as 
\begin{equation}
\mathring R = D^2 \Phi,
\label{Liouville}
\end{equation}
which implies 
\bea
D_A N_{\text{vac}}^{AB}=-\frac{1}{2}D^B \mathring R. \label{Li2}
\eea
We can therefore add a trace to $N^{\text{vac}}_{AB}$ to form the conserved stress-tensor
\begin{equation}
T_{AB}[\Phi]= \frac{1}{2}D_A \Phi D_B \Phi - D_A D_B \Phi + \frac{1}{2}q_{AB} \left( 2 D^2 \Phi  -\frac{1}{2} D^C \Phi D_C \Phi \right).
\end{equation}
Its trace is equal to $D^2 \Phi$. The tensor $T_{AB}$ is precisely the stress-tensor of Euclidean Liouville theory 
\bea
L[\Phi ; q_{AB}] = \sqrt{q}\left( \frac{1}{2} D^A \Phi D_A \Phi +\Lambda e^\Phi +\mathring R[q] \Phi  \right). 
\eea
where the parameter $\Lambda$ is zero in order to satisfy \eqref{Liouville}. Note that in order to derive the stress-tensor from the Lagrangian, one needs to set the Liouville field off-shell by not imposing the equation \eqref{Liouville} but considering the metric as a background field. Under a super-Lorentz transformation 
\bea
\delta_Y (D^2 \Phi - \mathring R) = (\mathcal L_Y + D_A Y^A)  (D^2 \Phi - \mathring R).\label{actionRL}
\eea 
Therefore, imposing the Liouville equation is consistent with the action of super-Lorentz transformations. 

Using this boundary metric and shear, one can work out the covariant expressions for $R_0$ and $R_1$ in \eqref{rc}. They are given by
\begin{equation}
R_0 = \frac{1}{2} e^{-\Phi}D^2 W \qquad \text{and} \qquad R_1 = \frac{1}{8} e^{-\Phi} C_{AB} C^{AB}.
\end{equation}
Finally, after some algebra, one can write the full metric as
\bea
ds^2 = - \frac{\mathring R}{2}\text{d}u^2 - 2 \text{d}\rho \text{d}u + (\rho^2 q_{AB} + \rho C^{\text{vac}}_{AB} + \frac{1}{8}C^{\text{vac}}_{CD}C_{\text{vac}}^{CD} q_{AB})\text{d}x^A \text{d}x^B + D^B C^{\text{vac}}_{AB} \text{d}x^A \text{d}u\label{metf}
\eea 
where $\rho = \sqrt{r^2+ \frac{1}{8}C^{\text{vac}}_{CD}C_{\text{vac}}^{CD}}$ is a derived quantity in terms of the Bondi radius $r$. The metric is more natural in Newman-Unti gauge $(u,\rho,z^A)$ where $g_{\rho \mu} = -\delta_\mu^\rho$. 

Let us also comment on the meromorphic extension of the Lorentz group instead of Diff$(S^2)$. When super-Lorentz transformations reduce to local conformal Killing vectors on $S^2$ \textit{i.e.} $G^z = G(z)$ and $G^{\bar z} \equiv \bar{G}(\bar z) $, the boundary metric after a diffeomorphism is the unit round metric on the sphere 
\bea
\bar{q}_{AB} \text{d}z^A \text{d}z^B = 2 \gamma_s \text{d}z \text{d}\bar{z},\qquad \gamma_s = \frac{2}{(1+z\bar z)^2}
\eea 
(and $\mathring R = 2$) \emph{except at the singular points of }$G(z)$. The Liouville field reduces to the sum of a meromorphic and an anti-meromorphic part minus the unit sphere factor 
\bea
\Phi = \phi(z)+\bar \phi(\bar z)- \log \gamma_s.\label{Phim}
\eea
The metric \eqref{metf} then exactly reproduces the expression of \cite{Compere:2016jwb} with the substitution $T^{(\text{there})}_{AB} = 1/2 N^{\text{vac}}_{AB}$. We have therefore found the generalization of the metric of the vacua for arbitrary $\text{Diff}(S^2)$ super-Lorentz transformations together with arbitrary supertranslations.

\subsection{The superboost, superrotation and supertranslation fields}

A general vacuum metric is parametrized by a boundary metric $q^{\text{vac}}_{AB}$, the field $C$ that we call the \emph{supertranslation field} and $\Phi$ that we will call either the \emph{Liouville field} or the \emph{superboost field}. Under a BMS transformation, the bulk metric transforms into itself, with the following transformation law of its boundary fields,
\bea
\delta_{T,Y}q^{\text{vac}}_{AB} &=& D_A Y_B + D_B Y_A - q^{\text{vac}}_{AB} D_C Y^C, \\
\delta_{T,Y}\Phi &=& Y^A \p_A \Phi + D_A Y^A,\\
\delta_{T,Y}C &=& T + Y^A \p_A C - \frac{1}{2}C D_A Y^A. \label{deltaC}
\eea
Only the divergence of a general super-Lorentz transformation sources the Liouville field. Since rotations are divergence-free but boosts are not, we call $\Phi$ the \emph{superboost field}. In general, one can decompose a vector on the 2-sphere as a divergence and a rotational part. For a generic superotation there should be a field that is sourced by the rotational of $Y^A$. We call this field the \emph{superrotation field} $\Psi$ and we postulate its transformation law
\bea
\delta_{T,Y}\Psi &=& Y^A \p_A \Psi + \eps^{AB} D_A Y_B.
\eea
Where is that field in \eqref{metf}? In fact, the boundary metric $q^{\text{vac}}_{AB}$ is not a fundamental field. It depends upon the Liouville field $\Phi$ and the background metric $\gamma_{ab}$. Since it transforms under superrotations, the metric \eqref{qABform} should also depend upon the superrotation field $\Psi$. The explicit form $q^{\text{vac}}_{AB}[\gamma_{ab},\Phi,\Psi]$ is not known to us. We will call the set of boundary fields $(\Phi,\Psi)$ the super-Lorentz fields.

Under a BMS transformation, the news of the vacua $N^{\text{vac}}_{AB}$ and  the tensor $C^{(0)}_{AB}$ transform inhomogenously as 
\bea\label{dRN}
\delta_{T,Y} N^{\text{vac}}_{AB} &=& \mathcal L_Y N^{\text{vac}}_{AB} -D_A D_B D_C Y^C + \frac{1}{2}q_{AB}D^2 D_C Y^C,\\\delta_{T,Y}C^{(0)}_{AB}  &=&  \mathcal L_Y C^{(0)}_{AB} - \frac{1}{2}D_C Y^C C^{(0)}_{AB}    -2 D_A D_B T + q_{AB}D^2 T.\label{baret}
\eea

From \eqref{metf}, one can read off the explicit expressions of the Bondi mass and angular momentum aspects of the vacua
\begin{equation}
\begin{split}
M &= -\frac{1}{8} N^{\text{vac}}_{AB} C_{\text{vac}}^{AB} ,\\
N_{A} &= -\frac{3}{32} D_A (C^{\text{vac}}_{BC}C_{\text{vac}}^{BC}) - \frac{1}{4} C^{\text{vac}}_{AB} D_C C_{\text{vac}}^{BC}. 
\end{split}\label{valvac}
\end{equation}
The Bondi mass is time-dependent and its spectrum is not bounded from below because $\p_u M = -\frac{1}{8}N_{AB}^{\text{vac}}N^{AB}_{\text{vac}}$ as observed in \cite{Compere:2016jwb}. Yet, the Weyl tensor is identically zero so the standard Newtonian potential vanishes. This indicates that the mass is identically zero. The relationship between the Bondi mass and the mass will be given below in Section \ref{sec:chv} after introducing a consistent phase space in which the conserved charges will be defined.

\section{Superboost transitions}
\label{sec:transitions}

The main interest of the non-trivial vacua lies in the dynamical processes that allow to transition from one vacuum to another. In what follows, we will relax the condition $\p_u q_{AB} = 0$ in order to allow for transitions of the super-Lorentz fields. We will study several examples of transitions and study the related memory effects at null infinity.

\subsection{Robinston-Trautman spacetimes}

The simplest example of spacetime containing a transition of the superboost field $\Phi$ is the general Robinson-Trautman spacetime\footnote{This metric is exactly (28.8) of \cite{2003esef.book.....S} with $P(u,\zeta,\bar\zeta)=e^{\Phi(u,\zeta,\bar \zeta)/2}$ after fixing the reparametrization ambiguity to set $M$ to a constant.}
\bea
\text ds^2 = -\Big(-  r \p_u \Phi +  \frac{\mathring R}{2} - \frac{2M}{r}\Big) \text{d} u^2 - 2 \text{d} u \text{d} r + 2 r^2 e^{-\Phi} \text{d}\zeta \text{d}\bar \zeta \label{RT}
\eea
where $\Phi=\Phi(u,\zeta,\bar\zeta)$ obeys the constraint
\bea
D^2 \mathring R + 12 M \p_u \Phi = 0. 
\eea
The Ricci scalar of the boundary metric is related to $\Phi$ by $\mathring R = D^2 \Phi$, which is the Liouville equation \eqref{Liouville}.

We consider a configuration where there is a transition between a perturbed vacuum at $u=u_i$ which relaxes to a new vacuum at $u = u_f$. The metric represents a transition from the Schwarzschild black hole equipped with an initial superboost field $\Phi(u_i,\zeta,\bar\zeta) = \Phi_i(\zeta,\bar\zeta) $ with $\p_u \Phi(u=u_i) \neq 0$ to the Schwarzschild black hole with a final superboost field $\Phi = \Phi_f(\zeta,\bar\zeta)$, $\p_u \Phi(u=u_f) \approx 0$. In other words, the Robinston-Trautman spacetime with $M \neq 0$ describes a gravitational wave emission process that evolves the Schwarzschild black hole with superboost hair.

The impulsive limit of the Robinson-Trautman type N of positive 2-curvature ($M= 0$, $\mathring R = 2$)  can be rewritten after a coordinate transformation as the metric of the impulsive gravitational waves of Penrose \cite{Penrose:1972aa,Nutku:1992aa} as shown in \cite{Podolsky:1999zr,Griffiths:2002gm}\footnote{It is exactly the solution (2.10) of \cite{Podolsky:2002sa} with $\eps = +1$ upon substituting $U \rightarrow u/\sqrt{2}$, $V \rightarrow -\sqrt{2}\rho$, $H \rightarrow -1/2 N^{\text{vac}}_{zz}$. Strictly speaking $g_{uu} = -1-\frac{D^2 \phi}{2}$ at the poles of the meromorphic function $\phi(z)$, but $g_{uu}=  -1$ otherwise.}
\bea
\text ds^2 = -\text{d}u^2 - 2 \text{d}\rho \text{d}u + \left[ \rho^2 q_{AB} + u \rho \Theta(u) N^{\text{vac}}_{AB} + \frac{u^2}{8}\Theta(u) N^{\text{vac}}_{CD}N_{\text{vac}}^{CD} q_{AB} \right] \text{d}x^A \text{d}x^B .\label{imp}
\eea
where $N^{\text{vac}}_{AB}  = \left[\frac{1}{2}D_A \phi_f D_B \phi_f - D_A D_B \phi_f  \right]^{\text{TF}}$. The vacuum news coincides with \eqref{PhiL} after substituting $\Phi = -\log\gamma_s + \phi_f$ as in \eqref{Phim}. This metric is in Newman-Unti gauge, not in Bondi gauge. It represents the transition between two vacua labelled by distinct meromorphic superboost fields\footnote{The singular impulsive limit requires to consider singular diffeomorphisms transitions which turn out to reduce to meromorphic superboost transitions.} (initial $\phi_i = 0$ for $u< 0$ and final $\phi_f = \phi(z)+\bar\phi(\bar z)$ for $u > 0$). The metric $q_{AB}$ is the unit sphere metric globally for $u<0$ and locally for $u > 0$ but it contains singularities at isolated points for $u > 0$. These singularities can be understood as a cosmic string decays \cite{Griffiths:2002hj,Griffiths:2002gm,Strominger:2016wns}. 

\subsection{General impulsive gravitational wave transitions}

In general, both the supertranslation field $C$ and the superboost field $\Phi$ can change with hard (finite energy) processes involving null radiation reaching $\mathcal I^+$. This null radiation can originate in matter or in gravity itself. Such processes induce vacuum transitions among initial $(C_-,\Phi_-)$ and final $(C_+,\Phi_+)$ boundary fields. The difference between these fields can be expressed in terms of components of the matter stress-tensor and metric potentials reaching $\mathcal I^+$. The simplest possible transition between vacua are shockwaves which carry a matter stress-tensor proportional to a $\delta(u)$ function, as in the original Penrose construction \cite{Penrose:1972aa}. A distinct vacuum lies on each side of the shockwave and the transition between the boundary fields is dictated by the matter stress-tensor. Such a general shockwave takes the form 
\bea
\text ds^2 = - \frac{\mathring R}{2}\text{d}u^2 - 2 \text{d}\rho \text{d}u + (\rho^2 q_{AB} + \rho C_{AB} + \frac{1}{8}C_{CD}C^{CD} q_{AB})\text{d}x^A \text{d}x^B + D^B C_{AB} \text{d}x^A \text{d}u\label{metf2}
\eea
where 
\bea
q_{AB}&=&\Theta(-u) q_{AB}^{\text{vac}}[\Phi_-] + \Theta(u) q_{AB}^{\text{vac}}[\Phi_+],\\
C_{AB}&=& \Theta(-u) C_{AB}^{\text{vac}}[\Phi_-, C_-] + \Theta(u) C_{AB}^{\text{vac}}[\Phi_+,C_+]
\eea
where $q_{AB}^{\text{vac}}[\Phi]$ and $C_{AB}^{\text{vac}}[\Phi,C]$ are given in \eqref{qABform} and \eqref{PhiL}. The metric \eqref{imp} is recovered for $\Phi_- = -\log\gamma_s$, $\Phi_+ = -\log\gamma_s + \phi(z)+\bar\phi(\bar z)$ as in \eqref{Phim} and $C_+ = C_- = 0$.

\subsection{Conservation of the Bondi mass aspect and the center-of-mass}

In the absence of superboost transitions and for the standard case of the unit round celestial sphere, the integral between initial $u_i$ and final retarded times $u_f$ of the conservation equation for the Bondi mass aspect \eqref{duM} can be reexpressed as the differential equation determining the difference between the supertranslation field $\Delta C = C_+ - C_-$ between initial and final retarded times after assuming suitable fall-off conditions \cite{Strominger:2014pwa}
\bea
-\frac{1}{4} D^2 (D^2 + 2)  \Delta C  =  \Delta M + \int_{u_-}^{u_+} \text{d}u \: T_{uu}\label{eq:98}
\eea
where $T_{uu}=\frac{1}{8}N_{AB}N^{AB}$ and $\Delta M$ is the difference between the Bondi mass aspects after and before the burst. The four lowest spherical harmonics $\ell=0,1$ are zero modes of the differential operator appearing on the left-hand side of \eqref{eq:98}. Recall that translations precisely shift the supertranslation field as \eqref{deltaC}. The 4 lowest harmonics of $C$ can thus be interpretated as the center-of-mass of the asymptotically flat system. This center-of-mass is not constrained by the conservation law \eqref{eq:98}.

A new feature arises in the presence of a superboost transition. The four zero modes of the supertranslation field $C$ are now determined by the conservation equation. This can be seen in the context of impulsive transitions \eqref{metf2}. For simplicity, we take $C_- = 0$ and $\Phi_- = -\log\gamma_s$ ($q_{AB}[\Phi_-] = \bar q_{AB}$ the unit round sphere metric). Given that the Bondi mass aspect and the Bondi news of the vacua are non-zero \eqref{valvac}, we first define the renormalized Bondi mass aspect and Bondi news as
\bea
\hat M &=& M + \frac{1}{8} C_{AB} N^{AB}_{\text{vac}}[\Phi_+],\\
\hat N_{AB} &=& N_{AB} - \Theta(u) N_{AB}^{\text{vac}}[\Phi_+],
\eea
which are zero for the vacua \eqref{metf}. This mass will be obtained in Section \ref{sec:phasespace} in \eqref{mf}. 

After integration over $u$ of \eqref{duM} and using of the corollary of the Liouville equation \eqref{Li2} we obtain 
\bea
-\frac{1}{4} D^2 (D^2 + \mathring R)  C_+ +\frac{1}{4}N^{AB}_{\text{vac}}[\Phi_+] D_A D_B C_+  + \frac{1}{8} C_+ D^2 \mathring R  = \Delta \hat M +  \int_{u_-}^{u_+} \text{d}u \: T_{uu} \label{eq4}
\eea
where $T_{uu}=\frac{1}{8}\hat N_{AB}\hat N^{AB}$ and $\Delta \hat M$ act as sources for $C_+$ and all quantities are evaluated on the final metric $q_{AB}[\Phi_+]$. We have that $\Delta \hat M=0$ for transitions between vacua but we included it for making the comparison with \eqref{eq:98} more manifest.

The lowest $\ell=0,1$ spherical harmonics of $C$ are not zero modes of the quartic differential operator on the left-hand side of \eqref{eq4} for any inhomogeously curved boundary metric. Therefore, the center-of-mass is also determined by the conservation law of the Bondi mass aspect.

\subsection{Refraction/Velocity kick memory}
%\label{sec:memory}

We will mostly consider the simplified case where the change of the boundary metric is localized at individual points. This happens for impulsive gravitational wave transitions which relate the initial and final boundary metric by a meromorphic super-Lorentz transformation (which is a combination of superboosts and superrotations). One example is the original Penrose construction \cite{Penrose:1972aa}. In these cases we will consider observers away from these singular points so that we can ignore these singularities. 

We can consider either timelike or null geodesics leading respectively to the velocity kick and refraction memory. Let us first discuss a congruence of timelike geodesics that evolve at finite large radius $r$ in the impulsive gravitational wave spacetime \eqref{imp}. Such observers have a velocity $v^\mu \p_\mu = \p_u + \mathcal{O}(\rho^{-1})$. The deviation vector $s^\mu$ between two neighboring geodesics obeys $\nabla_v \nabla_v s^\mu = R^\mu_{\;\; \alpha\beta\gamma}v^\alpha v^\beta s^\gamma$ where the directional derivative is defined as $\nabla_v = v^\mu \nabla_\mu$. We have $R_{uA uB}= -\frac{\rho}{2} \p^2_u C_{AB} + \mathcal{O}(\rho^0)$ where $C_{AB} = u \Theta(u) N_{AB}^{\text{vac}}$ and therefore 
\bea
q_{AB} \p_u^2 s^B &=& \frac{1}{2\rho} \delta(u) N_{AB}^{\text{vac}}s^B + \mathcal{O}(\rho^{-2}). 
\eea
We deduce that $s^A =s^A_{lead}(x^A)+ \frac{1}{\rho}s^A_{sub}(u,x^A)+\mathcal{O}(\rho^{-2})$ and after two integrations in $u$,  
\bea
s^A_{sub} = \frac{u}{2} \Theta(u) q^{AB } N_{BC}^{\text{vac}}s^C_{lead}. \label{devG}
\eea
Before the shockwave, there is no relative angular velocity between observers. After the shockwave, there will be a relative angular velocity at order $\propto \rho^{-1}$. This is the velocity kick between two such neighboring geodesics due to the shockwave \cite{Podolsky:2002sa,Podolsky:2010xh,Podolsky:2016mqg}. This is a qualitatively distinct effect from the displacement memory effect \cite{Zeldovich:1974aa,Blanchet:1987wq,Christodoulou:1991cr,Blanchet:1992br,0264-9381-9-6-018} and the spin effect \cite{Pasterski:2015tva}.

Analogously, one can consider a congruence of null geodesics which admits a constant leading angular velocity $\Omega^A(x^B)\p_A$, with total 4-velocity 
\bea
v^\mu \p_\mu = (\sqrt{\Omega^A q_{AB} \Omega^B} + \mathcal{O}(\rho^{-1}))\p_u + \mathcal{O}(\rho^{-1})\p_\rho + \frac{1}{\rho}( \Omega^A + \mathcal{O}(\rho^{-1}))\p_A . 
\eea
We consider again a deviation vector of the form $s^A =s^A_{lead}(x^A)+ \frac{1}{\rho}s^A_{sub}(u,x^A)+\mathcal{O}(\rho^{-2})$. The deviation vector obeys again \eqref{devG}. Null geodesics are refracted by the shockwave. This is the refraction memory effect usually described in the bulk of spacetime \cite{Podolsky:2002sa,Podolsky:2010xh,Podolsky:2016mqg}. We identified here the class of null geodesics which displays the refraction memory effect close to null infinity.  

Let us now shortly discuss the case where the change of boundary metric is not localized at individual points. This occurs in the example of Robinson-Trautman superboost transitions. The main point is that timelike geodesics will now admit non-trivial deviation vector already at leading order $\propto\, \rho^0$, $s^A =s^A_{lead}(x^A) +\mathcal{O}(\rho^{-1})$, with  
\bea
\frac{1}{2} q_{AB} \p_u^2 s_{lead}^B + \frac{1}{2} \p_u^2 (q_{AB} s^B_{lead}) &=&- \frac{1}{2} \p_u^2 q_{AB} s_{lead}^B . 
\eea
A velocity kick will therefore already occur at order $\rho^0$.

\subsection{A new non-linear displacement memory}
\label{sec:memory}

We also would like to point out that there is a non-linear displacement memory induced by a superboost transition, when it is accompanied by a supertranslation transition. This case was not considered in \cite{Podolsky:2002sa,Podolsky:2010xh,Podolsky:2016mqg} where all supertranslation transitions were vanishing. In order to describe the effect, we can consider either timelike or null geodesics. For definiteness, we consider a congruence of timelike geodesics that evolve at finite large radius $r$ in the general impulsive gravitational wave spacetime \eqref{metf2}. For simplicity we assume global Minkowski in the far past and we only consider the simplified case where the change of the boundary metric is localized at individual points. In other words, we assume $\Phi_- = -\log\gamma_s$ ($q^{\text{vac}}_{AB}[\Phi_-]$ is the unit sphere metric), $C_- = 0$, $\Phi_+ = -\log\gamma_s + \phi(z)+\bar \phi(\bar z)$ and $C_+=C_+(z,\bar z)$ arbitrary. The velocity is now $v^\mu \p_\mu = \sqrt{\frac{2}{\mathring R}}\p_u +\mathcal{O}(\rho^{-1})$. We have  $R_{uA uB}= -\frac{\rho}{2} \p^2_u C_{AB} + \mathcal{O}(\rho^0)$. Following the same procedure as above, we obtain  $s^A =s^A_{lead}(x^A)+ \frac{1}{\rho}s^A_{sub}(u,x^A)+\mathcal{O}(\rho^{-2})$ and away from the singular points on the sphere, 
\bea
s^A_{sub} &=& \frac{1}{2} q^{AB}C_{BC} s^C_{lead}. \\
&=& \frac{1}{2} \Theta(u)  q^{AB}C_{BC}^{\text{vac}} s^C_{lead}.\\
&=& \frac{1}{2} q^{AB} (u \Theta(u) N_{BC}^{\text{vac}} +  \Theta(u) C^{(0)}_{BC}  + \Theta(u) C N_{BC}^{\text{vac}}  )s^C_{lead}.
\eea
The first term $\propto\, u \Theta(u)$ leads to the velocity kick memory effect. The second term $\propto\, \Theta(u) C^{(0)}_{BC} $ leads to the displacement memory effect due to a change of supertranslation field $C$ between the final and initial states \cite{Strominger:2014pwa}.  The third and last term $\propto\, \Theta(u) C N^{\text{vac}}_{BC} $ is a new type of non-linear displacement memory effect due to change of both the superboost field $\Phi$ and the supertranslation field $C$. The four lowest spherical harmonics $\ell=0,1$ of $C$, interpretated as the center-of-mass, do not contribute to the standard displacement memory effect because they are zero modes of the differential operator $C_{AB}^{(0)}$. Here, they do contribute to the non-linear displacement memory effect. The transition  of the supertranslation field and in particular of the center-of-mass are determined by \eqref{eq4}, as discussed earlier.

\section{Renormalized phase space at $\mathcal I^+$}
\label{sec:phasespace}

In this section, we will define an extended phase space invariant under the action of Diff$(S^2)$ super-Lorentz transformations and supertranslations. Super-Lorentz transformations are overleading in the sense that they change the boundary metric which is usually fixed in standard asymptotically flat spacetimes. We can therefore expect that a renormalization procedure will be required.

\subsection{Boundary conditions}
\label{sec:BC2}

Following Campiglia and Laddha \cite{Campiglia:2015yka}, we fix the boundary metric determinant to be the one of the unit round sphere, 
\begin{equation}
\sqrt{q}  = \gamma_s, \qquad \gamma_s=2(1+z \bar z)^{-2}. \label{BC1}
\end{equation}

We use the fall-off conditions discussed in Section \ref{sec:BC1}. In addition, we impose the leading equations of motion \eqref{Ein2}, namely 
\bea
\mathring{V} = - \frac{1}{2} \mathring{R},\qquad \mathring{\beta} = -\frac{1}{32} C^{AB}C_{AB},\qquad \mathring{U}^A = -\frac{1}{2} D_B C^{AB}\label{BC3}
\eea 
as a part of the boundary conditions. 

The inhomogenous part of the transformation law of the news tensor under superboosts \eqref{dNAB} exactly matches with the transformation law of the vacuum news $N_{AB}^{\text{vac}}$ defined in \eqref{dRN}. Moreover, the inhomogenous part of transformation law of $C_{AB}^{\text{vac}}$ \eqref{PhiL} also matches with the one of $C_{AB}$ in \eqref{dCAB}. We are therefore led to 
introduce the initial $(C_-,\Phi_-,\Psi_-)$ and final $(C_+,\Phi_+,\Psi_+)$ boundary supertranslation and super-Lorentz fields and consider the following boundary conditions on the $q_{AB}$ and $C_{AB}$ tensors at $\mathcal I^+_\pm$,
\bea
q_{AB}^\pm  \equiv \lim_{u \rightarrow \pm\infty } q_{AB} &=& q^{\text{vac}}_{AB}[ \gamma_{ab}, \gamma_s ,\Phi_\pm ,\Psi_\pm] + \mathcal o(u^{0}), \label{BCC3}\\ 
\lim_{u \rightarrow \pm\infty } C_{AB} &=& C^{\text{vac}}_{AB}[q_{AB}^\pm, \Phi_\pm ,C_\pm ] + \mathcal o(u^{0}),\label{BCC2}
\eea
where $q_{AB}^{\text{vac}}$ and $C^{\text{vac}}_{AB}$ are defined in \eqref{qABform}, \eqref{PhiL}. It follows that
\bea
\lim_{u \rightarrow \pm\infty } N_{AB} = N^{\text{vac}}_{AB}[ q^\pm_{AB}, \Phi_\pm ] + \mathcal o(u^{-1}) \label{BCC}
\eea
where $N^{\text{vac}}_{AB}$ is defined in \eqref{PhiL}. The initial fields $(C_-,\Phi_-,\Psi_-)$ can change along $u$ with \emph{hard} (finite energy) processes involving null radiation reaching $\mathcal I^+$. In general this null radiation can originate from matter or from gravity itself. Here we restrict ourselves to gravity only. 

As a consequence of the boundary condition \eqref{BC1},
\bea
\p_u q_{AB}  = \partial_u(\log \sqrt{q}) q_{AB} =  0\qquad  \Rightarrow \qquad q^-_{AB} = q^+_{AB} \label{puqab}
\eea
prevents transitions between the initial and final superboost and superrotation fields $\Phi \equiv \Phi_+ = \Phi_-$, $\Psi \equiv \Psi_+ = \Psi_-$. The class of spacetimes that we are considering is therefore more general than the ones considered in \cite{Campiglia:2015yka,Christodoulou:1993uv} but not general enough to consider superboost transitions. We leave the contruction of a more general phase space for future endeavor.  

In addition, we impose that the topological Euler number of $q_{AB}$ is the one of the round sphere
\bea
\chi \equiv \frac{1}{4\pi} \int \sqrt{q}  \mathring{R}[q] = 2. \label{BC2}
\eea
Since this condition is diffeomorphic and Weyl invariant, it is consistent with the action of super-Lorentz transformations and supertranslations. We finally restrict the boundary metrics by imposing the Liouville equation
\begin{equation}
D^2 \Phi = \mathring R[q].
\label{Liouville equation boundary condition}
\end{equation} 
As shown in \eqref{actionRL}, imposing the Liouville equation is consistent with both the structure of the vacua and with the action of super-Lorentz transformations.

%This condition is automatically satisfied for a given solution if it is related by a super-Lorentz transformation to a solution for which the boundary metric is the unit round sphere. %In fact, in this case, the reasoning performed in section \ref{Generation of the vacua} for the vacua can be reproduced for this given solution. This means that the Christodoulou-Klainerman condition is violated only because of the presence of super-Lorentz transformations and can always be restored by choosing $q_{AB}$ to be the unit 2-sphere metric. 

The group of symmetries that preserve the boundary conditions are all the residual symmetries consisting of Diff$(S^2)$ super-Lorentz transformations and supertranslations. As we will see, after a suitable renormalization procedure, the canonical charges associated with these symmetries will be finite and non-trivial. The asymptotic symmetry group will therefore be the group of Diff$(S^2)$ super-Lorentz transformations and supertranslations. In what follows, we will discuss the action principle, the symplectic structure and the charges. All in all, these well-defined structures will allow to promote the solution space described in Section \ref{sec:Bondi} to a phase space, after imposing the boundary conditions.

\subsection{Examples of solutions}

%Let us first present some solutions obeying the boundary conditions. 

The Kerr black hole is obviously part of the phase space. Any vacuum gravitational field configuration that admits a wave-zone region and that does not contain incoming radiation is also part of this phase space since it admits a polynomial Bondi expansion \cite{Blanchet:1985sp}. These solutions have a trivial boundary metric, the unit round metric on $S^2$. 

An example of solution with non-trivial boundary metric is the following. The most general Robinson-Trautman metrics, \textit{i.e.} the general vacuum solution admitting a geodesic, shearfree, twistfree but diverging null congruence, are not part of the phase space because of our restriction \eqref{qu}. However, a subset of Robinson-Trautman metrics is part of the phase space. Let us start from (28.8) of \cite{2003esef.book.....S} with $P(\zeta,\bar \zeta)=e^{\Phi(\zeta,\bar \zeta)/2}$, 
\bea
\text ds^2 = -\left( \frac{D^2\Phi(\zeta,\bar \zeta) }{2} - \frac{2 M+ \frac{u}{4}D^2 D^2 \Phi(\zeta,\bar \zeta)}{r} \right) \text{d}u^2 -2 \text{d}u \text{d}r + 2 r^2 e^{-\Phi(\zeta,\bar \zeta) } \text{d}\zeta \text{d}\bar \zeta\label{RT2}.
\eea
Here $D^2 =D_A D^A = 2 e^{4\Phi}\p_z \p_{\bar z}$ and $\Phi(\zeta,\bar \zeta)$, $M$ is arbitrary. This is the Schwarzschild black hole dressed with a superboost field. Note that the Ricci scalar of the boundary metric is $\mathring R[q]=D^2 \Phi$. In order to obey the determinant condition \eqref{BC1} we need to consider a diffeomorphism $\zeta(z,\bar z)$, $\bar \zeta(z,\bar z)$ with $e^{-\Phi} (\p_z \zeta \p_{\bar z}\bar \zeta - \p_{\bar z} \zeta \p_z \bar \zeta)= \sqrt{\bar q} = 2(1+z \bar z)^{-2}$. The metric is then in Bondi gauge. One can write $\Phi = -\log \sqrt{\bar q} +\log (\p_z \zeta \p_{\bar z}\bar \zeta - \p_{\bar z} \zeta \p_z \bar \zeta)$. Since a diffeomorphism does not affect the topological condition \eqref{BC2}, one can evaluate it using $\zeta=z$, $\zeta=\bar z$ and check that it is obeyed using $D^2 ( -\log \sqrt{\bar q})=2$. The metric \eqref{RT2} in coordinates $(u,r,z,\bar z)$ is therefore part of the phase space. %Such metrics can be type II, D, III, N or O. 

\subsection{Action principle} 

\begin{figure}[h!]
\begin{center}
\begin{tikzpicture}[scale=0.8]
%\draw (0,0) circle (1pt);
%\draw[step=0.5,black!10,thin] (-2,-2) grid (4,4);
\draw[black] (0,4) node[above]{$\mathcal{I}^+_+$} -- (4,0) node[right]{$\mathcal{I}^+_-$};
\draw[blue] (-1.5,0.5) -- (0.5,2.5) -- (2,1) -- (0.0,-1);
\node[blue,rotate=-45] at (1.45,1.95) {$r=\Lambda$};
\node[blue,rotate=45] at (-0.75,1.75) {$u=u^+$};
\node[blue,rotate=45] at (-0.75,1.75) {$u=u^+$};
\node[blue,rotate=45] at (1.17,-0.17) {$u=u^-$};
\node[black,above] at (2.25,2.25) {$\mathcal{I}^+$};
\end{tikzpicture}
\end{center}\caption{Contour for the variational principle.}\label{FigV}
\end{figure}
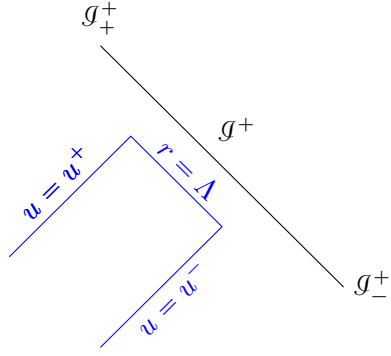

We consider the variation of the action on the spacetime volume $\mathcal M$ bounded in the following contour indicated in Figure \ref{FigV}. We denote as $\mathcal{I}_\Lambda^+$ and $\mathcal{U}^\pm$ the hypersurfaces $r=\Lambda$ and $u=u^\pm$, respectively. We consider the limit $\Lambda \rightarrow +\infty$, $u^\pm \rightarrow \pm\infty$.

The variation of the bare Einstein-Hilbert action is
\begin{equation}
\delta S_{EH} = \int_{\mathcal M} \text{d}u \, \text{d}r \, \text d^2 x \, \left\lbrace - \frac{\sqrt{-g}}{16\pi G} G^{\mn} \delta g_\mn + \partial_\mu \Theta^\mu [g,\delta g] \right\rbrace 
\end{equation}
where
\begin{eqnarray}
\Theta^u &=& r \,\Theta^u_{(div)} + \Theta^u_{(fin)} + r^{-1} \Theta^u_{(sub)} + \Order{2}\label{th1}, \\
\Theta^r &=& r \,\Theta^r_{(div)} + \Theta^r_{(fin)} + \Order{1}\label{th2}.
\end{eqnarray}
We have $\Theta^u_{(div)} \propto \delta \sqrt{q}$ and therefore  $\Theta^u_{(div)} =0$ as a result of the boundary condition \eqref{BC1}. Also, 
\bea
\Theta^u_{(sub)} &=& -8 \pref \delta \left[ \frac{1}{32} C_{AB}C^{AB} + \mathring{\beta}\right] =0
\eea
as a result of the boundary conditions \eqref{BC3}. The other components are
\begin{eqnarray}
\Theta^u_{(fin)} &=& \pref \frac{1}{2} C_{AB} \delta q^{AB}  ,\\
\Theta^r_{(div)} &=&2\pref \delta \mathring{V} - \frac{1}{2} \pref  N_{AB} \delta q^{AB} , \\
\Theta^r_{(fin)} &=& \pref \, \delta \left[ 2\partial_u \mathring{\beta} + 2 M + D_A \mathring{U}^A \right] + \bar \Theta_{flux} - \pref D_A (\mathring{U}_B \delta q^{AB} ), 
\end{eqnarray}
where we define with hindsight the important quantity
\bea
\bar \Theta_{flux} &\equiv&  \pref \Big[ \frac{1}{2}  N_{AB}\delta C^{AB} + \frac{1}{2}  \mathring{V} C_{AB} \delta q^{AB} +  \mathring{U}_B D_A \delta q^{AB} \Big].\label{thetaflux}
\eea

After using \eqref{Ein2}, we note that one can isolate a total derivative and a total variation as
\bea
\Theta^u_{(fin)} &=&- \p_r Y^{ur}, \label{d1}\\
r \Theta^r_{(div)}&=& -\p_u Y^{ru} + \delta (-\sqrt{q} \mathring{R}\; r) = -\p_u Y^{ru} -\p_A Y^{rA}\label{d2}
\eea
where $Y^{ur}=-Y^{ru}= -r \frac{1}{2} \pref C_{AB} \delta q^{AB}$ and $Y^{rA} = r \, \prefwg \, \Theta^A_{2d}(\delta q ; q)$ is $r$ times the presymplectic potential of the 2-dimensional Einstein-Hilbert action, $\p_A \Theta^A_{2d} =  \delta (\sqrt{q} \mathring{R})$. Since the boundary of a boundary is zero, the corner terms $\propto\, Y^{ur}$ in the variational principle drop. After integration over the sphere, the total derivatives $\propto\, \p_A v^A$ also drop. The radially divergent contribution to the action is therefore 
\bea
\prefwg \, \delta \left[ r \int_{u_i}^{u_f} \text du\, \chi[q] \right]
\eea
where $\chi[q]$ is the Euler number of the boundary metric. Under an infinitesimal smooth variation, a topological number cannot change. If we allow singular infinitesimal changes, such as the ones generated by singular super-Lorentz transformations that arise in the snapping of cosmic strings, the boundary topology changes and one would require to add a boundary term in the action to cancel this divergence.  Instead, this divergent term is zero using our boundary condition \eqref{BC2}.

We are led to consider the Einstein-Hilbert action supplemented by a boundary term
\bea
S = S_{EH} - \prefwg \int_{\mathcal I^+} \text{d}u \, \text{d}^2\Omega \, \Big( 2M- \frac{1}{8}C^{AB}N_{AB} \Big)
\eea
where $\text d^2\Omega \equiv \sqrt{q}\, \text d^2 x = \frac{1}{2}\sqrt{q}\, \epsilon_{AB}\text dx^A \text dx^B$. We do not provide a covariant formulation of this boundary term, or the boundary terms $Y^{\mu\nu}$, which would require geometrical tools on boundary null surfaces \cite{Parattu:2015gga,Lehner:2016vdi,Wieland:2017zkf,Duval:2014-2} or a prescription from holographic renormalization \cite{Kraus:1999di,Mann:2005yr}. The variation of the total action is 
\bea 
\delta S = \int_{\mathcal U^+} \Theta_{interior}^u - \int_{\mathcal U^-} \Theta_{interior}^u + \int_{\mathcal I^+_\Lambda} \bar \Theta_{flux}.  \label{deltaS}
\eea
where $\Theta_{interior}^u = \Theta^u -\Theta^u_{(fin)} = \mathcal O(r^{-2})$. All terms are radially finite and can be interpreted as follows. The spacetime  $\mathcal M$ considered is an open system with physical flux leaking through the surface $\mathcal I^+_\Lambda$. This leak needs to be exactly compensated by the difference between the fluxes on constant $u=u^+$ and $u = u^-$ slides in order to have a well-defined variational principle. Our analysis is insufficient to prove the existence of a variational principle but is compatible with its existence.

\subsection{Symplectic structure}

The bare Lee-Wald presymplectic form is given by $\omega_{(0)}[\delta_1 g, \delta_2 g] = \delta_1 \Theta_{(0)}[g,\delta_2 g] - \delta_2 \Theta_{(0)}[g,\delta_1 g]$. We already obtained that the bare presymplectic potential $\Theta_{(0)}$ is divergent while studying the variation of the action, see \eqref{th1}-\eqref{th2}. However, it is ambiguous under the change $\Theta_{(0)} \rightarrow \Theta_{(0)} + \text dY$ where $Y$ is a co-dimension spacetime 2 form. While studying the variational principle, we already identified in \eqref{d1}-\eqref{d2} the counterterms required to make the presymplectic potential finite.

Let us discuss this point in detail starting from the bare presymplectic form. We have 
\begin{align}
\omega_{(0)}^u &= \pref \left( \frac{1}{2} \delta_1 q_{AB} \delta_2 C^{AB} \right) + \mathcal{O}(r^{-2}) - (1 \leftrightarrow 2) ,\\
\omega_{(0)}^r &= r \pref \left( -\frac{1}{2} \delta_1 N_{AB} \delta_2 q^{AB} \right) \\
&\phantom{=} + \pref \left[ \frac{1}{2} \delta_1 \left( N^{AB} + \frac{1}{2} \mathring{R} q^{AB} \right)\delta_2 C_{AB} + \frac{1}{2} \delta_1 (D_A D^C C_{BC} ) \delta_2 q^{AB} \right] + \mathcal{O}(r^{-1}) - (1 \leftrightarrow 2). \nonumber
\end{align}
Clearly, such a presymplectic form is divergent. After choosing the boundary term $Y$ as  \eqref{d1}-\eqref{d2}, the presymplectic form becomes well-defined, 
\begin{align}
\omega^u &= \mathcal{O}(r^{-2}) - (1 \leftrightarrow 2) ,\\
\omega^r &= \pref \left[ \frac{1}{2} \delta_1 \left( N^{AB} + \frac{1}{2}\mathring{R} \, q^{AB} \right)\delta_2 C_{AB} + \frac{1}{2} \delta_1 (D_A D^C C_{BC}) \delta_2 q^{AB} \right] + \mathcal{O}(r^{-1}) - (1 \leftrightarrow 2). \label{eq:FiniteOmegaR}
\end{align}
(Note that since $Y^{Ar}$ is exact, it does not contribute here.) 
Since we specified a specific radial foliation in order to define these boundary terms, our construction is not explicitly covariant, but depends on additional background structures close to $\mathcal I^+$. We will not need to detail these boundary structures in the following. %However, we expect that the counterterm subtraction procedure that we used will lead to anomalies in the algebra of charges. It will be confirmed below, see \eqref{eq:Algebra}.  
This defines the symplectic structure at $\mathcal I^+$ 
\begin{eqnarray}
\Omega &=&  \frac{1}{16\pi G}\int_{\mathcal I^+} \text{d}u \, \text{d}^2 \Omega  \,\left[ \frac{1}{2} \delta_1 \left( N^{AB} + \frac{1}{2}\mathring{R} q^{AB} \right)\delta_2 C_{AB} + \frac{1}{2} \delta_1 (D_A D^C C_{BC}) \delta_2 q^{AB} \right]- (1 \leftrightarrow 2) \nonumber\\
&=& \int_{\mathcal I^+} \text{d}u \, \text{d}^2 x \left(  \delta_1 \bar \Theta_{flux}(\delta_2)  - \delta_2 \bar \Theta_{flux}(\delta_1) \right).\label{Omega}
\end{eqnarray}
where $\bar \Theta_{flux}$ is defined in \eqref{thetaflux}, after discarding a boundary term. 
%Yet, this symplectic structure admits divergences at $u \rightarrow \pm \infty$ when the boundary metric or the Liouville field vary. The renormalized symplectic structure is defined as 
%\begin{eqnarray}
%\Omega^{ren}&=& \frac{1}{16 \pi G} \int \text{d}u \, \text{d}^2 \Omega \left(  \delta_1 \Theta^{ren}_{flux}(\delta_2)  - \delta_2  \Theta^{ren}_{flux}(\delta_1) \right)\label{Omega}
%\end{eqnarray}
%where $\Theta^{ren}_{flux} = \bar\Theta_{flux} - u\, \bar\Theta_{flux}^{[1]} - \bar\Theta_{flux}^{[0]}- u^{-1} \bar\Theta_{flux}^{[-1]} $ as defined in \eqref{bndthetabar}.
\subsection{Surface charges} 

\subsubsection{Infinitesimal surface charges}  The bare Iyer-Wald  surface charge $2-$form is
\begin{align}
k [g,\delta g] = -\delta Q_\xi [g] + Q_{\delta\xi} [g] + i_\xi \Theta_{(0)} [g,\delta g].\label{decom:charge}
\end{align}
Expanding the $ur$ component in powers of $1/r$, we get
\begin{equation}
k^{ur} [g,\delta g] = r k^{ur}_{(div)} + k^{ur}_{(fin)} + \mathcal{O}(r^{-1}).
\end{equation}
We define $\slashed{\delta}\bar H_\xi = \oint k_\xi [g,\delta g] = \oint \text d^2 x \, k^{ur}_\xi[g,\delta g]$, where the integration $\oint$ is performed on the celestial sphere $S^2_\infty$. The divergent term is
\begin{equation}
\slashed{\delta}\bar H_\xi^{(div)} = \prefwg \oint \text{d}^2 \Omega \left[ -2 \delta(Y^A \mathring{U}_A) - f \delta \mathring{R} - \frac{1}{2} f N_{AB} \delta q^{AB} + \frac{1}{4} D_C Y^C q_{AB} \delta C^{AB} \right]\label{divch}
\end{equation}
while the finite term is
\begin{equation} 
\begin{split}
\slashed{\delta} \bar H^{(fin)}_{\xi} &= \frac{1}{16\pi G} \oint \text{d}^2\Omega \left[ 4 f M + 2 Y^A N_A + \frac{1}{16} Y^A \partial_A (C_{BC}C^{BC}) \right]\\
&\quad + \frac{1}{16\pi G} \oint \text{d}^2\Omega \left[ \frac{1}{2}f \left(N^{AB}+\frac{1}{2}q^{AB} \mathring R \right) \delta C_{AB}- 2\partial_{(A} f \mathring U_{B)} \delta q^{AB}\right. \\
&\qquad \qquad \qquad \qquad \left. - f D_{(A} \mathring U_{B)}\delta q^{AB} -\frac{1}{4}D_C D^C f C_{AB}\delta q^{AB}\right].
\end{split}
\label{Hxifin}
\end{equation}
Clearly, the charges are neither finite nor integrable. 

Incorporating the boundary counterterm $Y$ as defined  in \eqref{d1}-\eqref{d2}, the charges become
\begin{equation}
k^{ur}_\xi \rightarrow k^{ur}_\xi - \delta Y^{ur}[g, \mathcal{L}_\xi g ] + Y^{ur}[g, \mathcal{L}_{\delta\xi} g] + (\xi^u \Delta \Theta^r - \xi^r\vert_{\text{leading}} \Delta \Theta^u).\label{shiftk}
\end{equation}
In details,
\begin{align}
- \delta Y^{ur}[g, \mathcal{L}_\xi g ] &= \frac{\sqrt{q}}{16\pi G} r \delta \left( C_{AB} \delta_{(T,Y)} q^{AB} \right) \label{1} ;\\
Y^{ur}[g, \mathcal{L}_{\delta\xi} g] &=0 \label{12},
\end{align}
where the last equation follows from the fact that the fields are not modified by $\delta \xi$ at leading order in $r$. Moreover, 
\begin{align}
\xi^u \Delta \Theta^r &= f (\partial_u Y^{ur} + \partial_A Y^{Ar}) = \frac{1}{2} r \frac{\sqrt{q}}{16\pi G} f N_{AB} \delta q^{AB} + \frac{\sqrt{q}}{16\pi G} r \left( f \delta \mathring{R}  \right) ;\label{2} \\
- \xi^r\vert_{\text{leading}} \Delta \Theta^u &= -\xi^r\vert_{\text{leading}}(\partial_r Y^{ru} ) = -\frac{1}{4} \frac{\sqrt{q}}{16\pi G} r D_C Y^C C_{AB} \delta q^{AB} . \label{3}
\end{align}
The prescription for using $\xi^r \vert_{\text{leading}}$ instead of $\xi^r$ is justified as follows. The incorporation of $Y$ renormalizes the symplectic current, which leads to the finite expression of $\omega^r$ in \eqref{eq:FiniteOmegaR}. The associated renormalized surface charges are obtained through the fundamental relation of the Iyer-Wald formalism \cite{Iyer:1994ys,Compere:2018aar}
$
\partial_u \left( \slashed{\delta}H_\xi[g,\delta g] \right) = \oint \text d^2 x\, \omega^r[g,\mathcal L_\xi g,\delta g ].
$ 
By construction, any divergent term will disappear in the infinitesimal surface charges thanks to our choice of $Y$ without modifiying the finite piece. Hence the final result for the infinitesimal surface charges reads as
\begin{equation} 
\ndelta H_{\xi} = \slashed{\delta} \bar H^{(fin)}_{\xi} . 
\label{infinitesimal charges}
\end{equation}

When $q_{AB}$ is the fixed unit metric on the sphere, it reproduces the expression of Barnich-Troessaert \cite{Barnich:2011mi}. One can split $\slashed{\delta}H_\xi$ into integrable and non-integrable parts as
\begin{equation}
\slashed{\delta}H_\xi [g,\delta g] = \delta H^{int}_\xi [g] + \Xi_\xi [g, \delta g].
\label{decomposition charge}
\end{equation} 
Here we defined the integrated charge as in Barnich-Troessaert \cite{Barnich:2011mi} or Flanagan-Nichols \cite{Flanagan:2015pxa} 
\bea
H^{int}_\xi [g] = \prefwg \oint \text{d}^2 \Omega  \left( 4 f M + 2 Y^A N_A + \frac{1}{16} Y^A D_A (C_{BC}C^{BC}) \right) \label{FNCharges}
\eea
but where $Y^A$ is now an arbitrary vector. The non-integrable piece reads as
\begin{equation}
\begin{split}
\Xi_\xi[g,\delta g] &= \frac{1}{16\pi G} \oint \text{d}^2\Omega \left[ \frac{1}{2}f \left(N^{AB}+\frac{1}{2}q^{AB} \mathring R \right) \delta C_{AB}- 2\partial_{(A} f \mathring U_{B)} \delta q^{AB}\right. \\
&\qquad \qquad \qquad \qquad \left. - f D_{(A} \mathring U_{B)}\delta q^{AB} -\frac{1}{4}D_C D^C f C_{AB}\delta q^{AB}\right].
\end{split}
\end{equation}

Of course, the canonical Hamiltonian cannot be deduced solely from the relation \eqref{decomposition charge} since one can shift $H_\xi$ as
\begin{equation}
\slashed{\delta}H_\xi [g,\delta g] = \delta (H_\xi [g] +\Delta H_\xi[g])  + \Xi_\xi [g, \delta g] - \delta \Delta H_\xi[g].\label{shifts}
\end{equation} 
We therefore need additional input to fix the finite Hamiltonian.

Moreover, the infinitesimal charges are still divergent in $u$. This divergence can be rooted to the action principle.  In \eqref{deltaS}, $\bar \Theta_{flux} = \mathcal O(u)$ which leads to divergences at $\mathcal I^+_\pm$ that require further renormalization. In principle, these can be absorbed by including the contribution of radially finite boundary counterterms $Y^{ur}$ which were left unfixed so far. Such additional boundary counterterms also regularize the $u$ divergences of the symplectic structure \eqref{Omega}. This procedure amounts to shift both $H^{int}_\xi[g]$ and $\Xi_\xi [g, \delta g]$ with a priori distinct contributions that depend upon the boundary fields defined at $\mathcal I^+_-$, $(C_-, \Phi_-, q^-_{AB})$. This counterterm subtraction therefore is more general than the shifts \eqref{shifts}. In this procedure, there remains a finite ambiguity due to the finite counterterm $Y^{ur}_{ambiguity}(x^A)$ that only depends upon the boundary fields $(C_-,\Phi_-,q_{AB}^-)$. This remaining ambiguity takes a very specific form since according to \cite{Iyer:1994ys} it shifts the surface charge as 
\bea
\slashed{\delta}H_\xi  \mapsto \slashed{\delta}H_\xi  - \delta \oint \text{d}^2 \Omega \left( Y^{ur}_{ambiguity}[C_-,\Phi_-,q_{AB}^- ; \delta_{(T,Y)}C_-,\delta_{(T,Y)}\Phi_-,\delta_{(T,Y)} q_{AB}^-] \right). \label{amb}
\eea
General theorems on the uniqueness of conserved charges in gravity are insufficient to remove such an ambiguity \cite{Barnich:1994db,Barnich:2001jy}. Only for exact Killing vectors this ambiguity vanishes. In summary, covariant phase space methods only fix the infinitesimal charge variation, not the finite charge variation due to the lack of integrability and, moreover, the counterterm subtraction procedure suffers from an ambiguity \eqref{amb}. A prescription is therefore required to define the finite charge associated with the asymptotic symmetries.

\subsubsection{Finite charge}

In order to define the finite charges $H_\xi$ associated with $\xi$ we will follow a more direct route which we will justify by its consistency with the soft theorems, the action of asymptotic symmetries and the vanishing energy of the vacua. 
Following the Wald-Zoupas procedure \cite{Wald:1999wa}, it would be natural to request that the flux $\p_u H_\xi [g] $ is identically zero in the absence of news. However, the news tensor transforms inhomogeneously under (both Virasoro and Diff$(S^2)$) super-Lorentz transformations so this condition is not invariant under the action of the asymptotic symmetry group. Instead, we request that the flux $\p_u H_\xi [g] $ is identically zero in the absence of shifted news $\hat N_{AB}$, 
\bea
\hat N_{AB} = N_{AB} - N_{AB}^{\text{vac}}[\Phi_-]. \label{cond1}
\eea
Since the latter transforms homogeneously under super-Lorentz transformations, this prescription is \textit{invariant} under the action of all asymptotic symmetries\footnote{Our definition of the invariant news tensor is motivated from the structure of the vacua and differs from the one of \cite{Ashtekar:1978zz,Ashtekar:2014zsa} defined from the ``$2d$ Weyl'' tensor of Geroch  \cite{1977asst.conf....1G}.}. For future use, we define the shifted $\hat C_{AB}$ tensor
\bea
\hat C_{AB} = C_{AB} - u N_{AB}^{\text{vac}}[\Phi_-]. \label{cond2}
\eea
such that $\p_u \hat C_{AB} = \hat N_{AB}$. In order to obtain our ansatz, let us start with the charge \eqref{FNCharges}. The flux associated to \eqref{FNCharges} reads as
\begin{align}
\partial_u H^{int}_\xi [g] = -\frac{1}{32\pi G} \oint \text{d}^2 \Omega &\left[  f N_{AB} N^{AB} -2 f D_A D_B N^{AB} - f D_A D^A \mathring{R}  - Y^A\mathcal{H}_A (N,C) \right. \\
&\left.  \quad + Y^A D_B D^B D^C C_{AC} - Y^A D_B D_A D_C C^{BC} - Y^A C_{AB} D^B \mathring{R} \right] .\nonumber
\end{align}
Here we defined for later convenience the bilinear operator on rank 2 spherical traceless tensors $P_{AB}$ and $Q_{AB}$:
\begin{equation}
\mathcal{H}_A (P,Q) \equiv \frac{1}{2} \partial_A (P_{BC} Q^{BC}) - P^{BC} D_A Q_{BC} + D_B (P^{BC} Q_{AC} - Q^{BC} P_{AC})
\end{equation}
which enjoys the property $\mathcal{H}_A(P,P)=0$. When $\hat N_{AB} = 0$ we are left with
\begin{equation}
\begin{split}
\p_u H^{int}_\xi |_{\hat N_{AB}=0} = -\frac{1}{32\pi G}\oint \text{d}^2\Omega &\left[ f N_{AB}^{\text{vac}} N^{AB}_{\text{vac}} - Y^A\mathcal{H}_A (N^{\text{vac}},C) - Y^A C_{AB} D^B \mathring{R} \right.   \\
&\left. \quad + Y^A D_B D^B D^C C_{AC} - Y^A D_B D_A D_C C^{BC} \right]
\end{split}
\label{eq:FluxNABphys}
\end{equation}
after using the relation \eqref{Li2} which follows from \eqref{Liouville equation boundary condition}.
We now want to define a counterterm that is only built out of the fields at $\mathcal I^+$ ($q_{AB},C_{AB},N_{AB}$) and out of $N_{AB}^{\text{vac}}$, which is the only boundary field that appears in the condition \eqref{cond1}. Our prescription that cancels the right-hand side of \eqref{eq:FluxNABphys} is 
\bea
\Delta H_\xi[g ; \Phi] &\equiv& \prefwg \oint \text{d}^2 \Omega \Big[ \frac{u}{2} Y^A D_B D^B D^C C_{AC} - \frac{u}{2} Y^A D_B D_A D_C C^{BC}   - \frac{u}{2} Y^A C_{AB} D^B \mathring{R} \nonumber \\
  &&\hspace{-1.5cm}+ \frac{1}{2} T C_{AB} N_{\text{vac}}^{AB} - \frac{u}{2} Y^A\mathcal{H}_A(N^{\text{vac}},C) + \frac{u^2}{8} D_C Y^C N^{\text{vac}}_{AB} N_{\text{vac}}^{AB} + \frac{u^2}{4} Y^A N^{\text{vac}}_{AB} D^B \mathring{R} \Big].
\label{DeltaH}
\eea
There is a considerable ambiguity in defining this ansatz since we could add terms of the form $\hat N_{AB} \mathcal{A}^{AB}(q,C,N^{\text{vac}})+\mathcal F(q,N^{\text{vac}})+\hat C_{AB}\mathcal{B}^{AB}(q,N^{\text{vac}})$ where $\mathcal{A}^{AB}$, $\mathcal F$ and $\mathcal{B}^{AB}$ are arbitrary functions linear in either $T$ or $Y$. We will justify our ansatz by showing consistency with the leading and subleading soft theorems, and consistency for defining the charges of the vacua. 

Our final prescription for the canonical charges is $H_\xi [g] = H^{int}_\xi [g] +\Delta H_\xi[g]$. The charges are conveniently written as
\begin{equation}
 \boxed{H_\xi[g] = \frac{1}{16 \pi G}\oint \text{d}^2 \Omega \left[  4 T \hat M  + 2 Y^A  \hat N_A \right]}\label{Hhat}
\end{equation}
where the final mass and angular momentum aspects are given by 
\bea
\hat M &=& M + \frac{1}{8} C_{AB} N^{AB}_{\text{vac}}, \label{mf}\\
\hat N_A &=& N_A -u \p_A M + \frac{1}{32} \partial_A (\hat C_{CD}\hat C^{CD}) + \frac{u}{16} \p_A (\hat C^{CD} N_{CD}^{\text{vac}})  - \frac{1}{32} u^2 \p_A (N_{BC}^{\text{vac}} N^{BC}_{\text{vac}}) \label{NAf}\\
&& - \frac{u}{4} \mathcal{H}_A (N^{\text{vac}},\hat C) - \frac{u}{4} \hat C_{AB} D^B \mathring{R}  
+ \frac{u}{4} D_B D^B D^C \hat C_{AC} - \frac{u}{4} D_B D_A D_C \hat C^{BC} -  \frac{u^2}{8} N_{AB}^{\text{vac}} D^B \mathring{R}.\nonumber
\eea
This is a new prescription for the charges. In the standard asymptotically flat spacetimes where the boundary metric is the round sphere ($q_{AB} = \bar q_{AB}$ with $\mathring R =2$), our expressions reduce to 
\begin{equation}
\begin{split}
\hat M &= M, \\
\hat N_A &= N_A -u \p_A M + \frac{1}{32} \partial_A (C_{CD} C^{CD})+ \frac{u}{4} D_B D^B D^C C_{AC} - \frac{u}{4} D_B D_A D_C C^{BC}.\label{finalNA}
\end{split}
\end{equation}
The Lorentz charges differ from the existing prescriptions \cite{Barnich:2011mi,Flanagan:2015pxa,Hawking:2016sgy} since the angular momentum aspect is now enhanced with the two soft terms linear in $u$. We will show that our prescription correctly reproduces the fluxes needed for the subleading soft theorem. 

\subsection{Flux formulae for the soft Ward identities}

Let us show that our expressions for the fluxes reproduce the expressions of the literature used in the Ward identities displaying the equivalence to the leading \cite{Weinberg:1965nx} and subleading \cite{Cachazo:2014fwa} soft graviton theorems. The final flux can be decomposed in soft and hard parts, where the soft terms (resp. hard terms) are linear (resp. quadratic) in $\hat C_{AB}$ or its time variation $\hat N_{AB}$. We have
\begin{equation}
\int_{\mathcal I^+} \text{d}u \, \p_u H_\xi[g] = Q_S[T] + Q_H[T] + Q_S[Y] + Q_H[Y]
\label{eq:FinalFlux}
\end{equation}
where 
\begin{align}
Q_S[T] &= \frac{1}{16\pi G}\int_{\mathcal I^+} \text{d}u \, \text{d}^2\Omega\; \p_u \left( T D_A D_B \hat C^{AB} \right), \label{SoftT} \\
Q_H[T] &= \frac{1}{16\pi G}\int_{\mathcal I^+} \text{d}u \, \text{d}^2\Omega \left( - \frac{1}{2} T \hat N_{AB}  N^{AB}  \right), \label{HardT} \\
Q_S[Y] &= \frac{1}{16\pi G}\int_{\mathcal I^+} \text{d}u \, \text{d}^2\Omega \;u \,\p_u \left(  \hat C^{AB} s_{AB} \right), \label{SoftY} \\
Q_H[Y] &= \frac{1}{16\pi G}\int_{\mathcal I^+} \text{d}u \, \text{d}^2\Omega \left(\frac{1}{2} Y^A \mathcal{H}_A(\hat N ,\hat C) + \frac{u}{2} Y^A N^C_D D_A \hat N^D_C + \frac{u}{2}  N^{CD}_{\text{vac}} Y^A D_A \hat N_{CD} \right) \label{HardY}
\end{align}
and 
\begin{equation}
s_{AB} = \left[  D_A D_B D_C Y^C + \frac{\mathring{R}}{2} D_{(A} Y_{B)} - \frac{1}{2}D_{(A} \Big(D^2 + \frac{\mathring{R}}{2}\Big) Y_{B)} \right]^{\text{TF}}
\label{sAB}
\end{equation}
after integrations by parts on the sphere. 

In the standard case where $N_{AB}^{\text{vac}} = 0$, the flux of supermomenta reproduces (2.11) of \cite{He:2014laa} up to a conventional overall sign, which itself agrees with previous results \cite{Ashtekar:1978zz}. After one imposes the antipodal matching condition on $\hat M$ at spatial infinity, one can equate the flux on $\mathcal I^+$ with the antipodally related flux on $\mathcal I^-$. The result of \cite{He:2014laa} is precisely that the quantum version of this identity is the Ward identity of the leading soft graviton theorem. We have now obtained a generalization in the presence of superboost background flux.

%We now consider the hard terms for super-Lorentz transformations. Using the identities
%\begin{equation}
%\begin{split}
%D_A \hat C_{BC} \hat  N^{BC} &= D_B \hat C_{CA}\hat N^{BC} + \hat N_{AB} D_C \hat C^{BC} ,\\
%D_A \hat N_{BC} \hat C^{BC} &= D_B \hat N_{CA} \hat C^{BC} + \hat C_{AB} D_C \hat N^{BC},
%\end{split}
%\end{equation} and integrating by parts, it can be shown that \eqref{HardY} can be rewritten as 
We now consider the hard terms for super-Lorentz transformations. Using the identity
\begin{equation}
\frac{1}{2} ( \hat N^{AC} \hat C_{BC} + \hat C^{AB} \hat N_{BC}) = \frac{1}{2} (\hat N^{BC} \hat C_{BC}) \delta^A_B
\end{equation} and integrating by parts, it can be shown that \eqref{HardY} can be rewritten as 
\begin{align}
Q_H[Y] &= \prefwg \int_{\mathcal I^+} \text{d}u \, \text{d}^2\Omega \, \left[- \frac{1}{2} \hat N^{AB} \left(\mathcal{L}_Y \hat C_{AB} - \frac{1}{2} D_C Y^C \hat C_{AB} + \frac{u}{2} D_C Y^C \hat N_{AB} \right) + u N^{AB}_{\text{vac}} Y^C D_C \hat N_{AB} \right] \nonumber \\
&= -\frac{1}{32 \pi  G}\int_{\mathcal I^+} \text{d}u\, \text{d}^2\Omega \, \left[ \hat N^{AB} \delta^H_Y \hat C_{AB} - 2u N^{AB}_{\text{vac}} Y^C D_C \hat N_{AB} \right]
\end{align}
where $\delta^H_Y$ is the homogeneous part of the transformation of $\hat C_{AB}$. After restricting to standard configurations where $N_{AB}^{\text{vac}} = 0$, the expression matches (up to the overall conventional sign) with equation (40) of \cite{Campiglia:2015yka}. 

Next, we consider the soft terms for super-Lorentz transformations. Noting that $D^C \delta_Y q_{AC} = D_C D^C Y_A + \frac{\mathring{R}}{2} Y_A$ we can rewrite \eqref{sAB} as
\begin{equation}
s_{AB} = \left[  D_A D_B D_C Y^C + \frac{\mathring{R}}{2} D_{(A} Y_{B)} - \frac{1}{2}D_{(A} D^C \delta_Y q_{B)C} \right]^{\text{TF}}.
\end{equation}
The tensor $s_{AB}$ is recognized as the generalization of equation (47) of \cite{Campiglia:2015yka} in the presence of non-trivial boundary curvature. After some algebra, we can rewrite it in terms of the inhomogeous part $\delta^I_Y C_{AB}$ of the transformation law of $C_{AB}$ \eqref{dCAB}:
 \bea
-u s_{AB} = \delta_Y^I C_{AB} \equiv -u (D_A D_B D_C Y^C + \frac{1}{2}q_{AB} D_C D^C D_E Y^E). 
\eea
Now that we identified our expressions with the ones of \cite{Campiglia:2015yka}, we can use their results. After imposing the antipodal matching condition on $\hat N_A$ at spatial infinity, one can equate the flux of super-Lorentz charge on $\mathcal I^+$ with the antipodally related flux on $\mathcal I^-$ as originally proposed in \cite{Hawking:2016sgy} (but where the expression for $\hat N_A$ should be modified to \eqref{finalNA}). The result of \cite{Campiglia:2015yka} is precisely that this identity is the Ward identity of the subleading soft graviton theorem \cite{Cachazo:2014fwa}.

We end up with two further comments. Note that the soft charges for super-Lorentz transformations agree with equation (41) of \cite{Campiglia:2015yka} (up to an overall conventional sign) after a partial integration on $u$ \textit{and} after using the restrictive boundary condition $\hat C_{AB} = \mathcal{o}(u^{-1})$,
\begin{equation}
\begin{split}
\pref \int_{\mathcal I^+} \text{d}u \, \hat C^{AB}  s_{AB} &= \pref \left[u \hat C^{AB} s_{AB}\right]_{\mathcal{I}_-^+}^{\mathcal{I}_+^+} - \pref \int_{\mathcal I^+} \text{d}u \, (u \hat N^{AB}  s_{AB})\\
&= - \pref \int_{\mathcal I^+} \text{d}u \, (u \hat N^{AB} s_{AB}) = -Q_S [Y].
\end{split}
\end{equation}
However, the boundary condition $\hat C_{AB} = \mathcal{o}(u^{-1})$ is not justified since displacement memory effects lead to a shift of $C$, \textit{e.g.} in binary black hole mergers. Therefore, using the more general boundary conditions, the valid expression for the soft charge is only given by \eqref{SoftY}. 

Finally, considering only the background Minkowski spacetime ($q_{AB} =\bar {q}_{AB}$ the unit metric on the 2-sphere and $N_{AB}^{\text{vac}} = 0$), one can check that in stereographic coordinates one has $s_{zz} =  \partial_z^3 Y^z =  D_z^3 Y^z$. The soft charge then reads as 
\begin{equation}
Q_S [Y] = \frac{1}{16\pi G}\int_{\mathcal I^+} \text{d}u \, \oint \text{d}^2 z \, \gamma_{s} \, (u \hat N^{zz} D_{{z}}^3 Y^{{z}} + u \hat N^{\bar{z}\bar{z}} D_{\bar{z}}^3 Y^{\bar{z}} )
\end{equation}
where we keep $Y^A \partial_A = Y^z (z,\bar{z}) \partial_z + Y^{\bar{z}} (z,\bar{z}) \partial_{\bar{z}}$ arbitrary. In the case of meromorphic super-Lorentz tranformations, this reproduces equation (5.3.17) of \cite{Strominger:2017zoo} (up to a conventional global sign).  This concludes our checks with the literature. It shows that the Ward identities of supertranslations and super-Lorentz transformations are equivalent to the leading and subleading soft graviton theorems following the arguments of \cite{He:2014laa,Campiglia:2014yka}.

\subsection{Charges of the vacua}
\label{sec:chv}

Using the values \eqref{valvac} in our prescription \eqref{Hhat} we deduce the mass and angular momenta of the vacua
\begin{equation}
H^{\text{vac}}_\xi [\Phi,C]= \frac{1}{16\pi G} \oint \text{d}^2\Omega \left[4 T \hat{M}^{\text{vac}} +  2 Y^A \hat N_A^{\text{vac}} \right]
\end{equation} where
\begin{equation}
\hat{M}^{\text{vac}} =0, \quad \hat N_A^{\text{vac}} = -\frac{1}{4} \hat C_{AB} D_C \hat C^{BC} - \frac{1}{16} \partial_A (\hat C_{CD} \hat C^{CD}),
\end{equation}
and $\hat C_{AB}= C N_{AB}^{\text{vac}} - 2 D_A D_B C + q_{AB} D^C D_C C$ in this case. 

The supermomenta are all identically vanishing. Remember that the Lorentz generators are uniquely defined as the 6 global solutions $Y^A$ to the conformal Killing equation $D_A Y_B + D_B Y_A = q_{AB} D_C Y^C$. 
%, which implies 
%\bea
%(D_A D_B + q_{AB})D_E Y^E = 0, \qquad (D^2 + 2)D_A Y^A = 0,\qquad (D^2 +1 )Y^A = 0. 
%\eea
In general, the Lorentz charges as well as the super-Lorentz charges are non-vanishing. 

For the round sphere metric $q_{AB} = \bar q_{AB}$ ($\Phi = -\log \gamma_s$), we have $\mathring{R}=2$, $N_{AB}^{\text{vac}} = 0$ and $D^B \hat C_{AB} = D^B C_{AB}^{(0)} =-D_A (D^2+2)C$. The charges then reduce to 
\begin{equation}
H^{\text{vac}}_\xi[C] = \frac{1}{8\pi G} \oint \text{d}^2\Omega \left[ T \times 0 +  Y^A \Big( -\frac{1}{4} C^{(0)}_{AB} D_C C_{(0)}^{BC} - \frac{1}{16} \partial_A (C_{CD}^{(0)} C^{CD}_{(0)})\Big)\right].
\end{equation} 
As shown in the appendix A.3 of \cite{Compere:2016jwb}, the Lorentz charges are identically zero. The difference of charges between our prescription and the one of \cite{Compere:2016jwb} are the last two terms of \eqref{finalNA} which exactly cancel for the vacua with a round sphere boundary metric. Therefore, we confirm that the vacua with only the supertranslation field turned on do not carry Lorentz charges. The super-Lorentz charges are conserved and non-vanishing in general, which allows to distinguish the vacua.

\subsection{Charge algebra}

After an involved computation, we get the following charge algebra
\begin{equation}
\boxed{
\delta_{\xi_1}H_{\xi_2}[g] + \Xi_{\xi_1}[g,\delta_{\xi_2}g] = H_{[\xi_2,\xi_1]}[g]+\mathcal K_{\xi_1,\xi_2}[g].
}
\label{eq:Algebra}
\end{equation}
In this relation,
\begin{equation} 
\begin{split}
\Xi_{\xi_1}[g,\delta_{\xi_2}g] &= \frac{1}{16\pi G} \oint \text{d}^2\Omega \left[ \frac{1}{2}f_1 \left(N^{AB}+\frac{1}{2}q^{AB} \mathring R \right) \delta_{\xi_2} C_{AB}- 2\partial_{(A} f_1 \mathring U_{B)} \delta_{\xi_2} q^{AB}\right. \\
&\qquad \qquad \qquad \quad\, \left. - f_1 D_{(A} \mathring U_{B)}\delta_{\xi_2} q^{AB} -\frac{1}{4}D_C D^C f_1 C_{AB}\delta_{\xi_2} q^{AB}\right]  - \delta_{\xi_2}\Delta H_{\xi_1},
\end{split}
\label{eq:XixiEval}
\end{equation}
and the 2-cocycle $\mathcal{K}_{\xi_1,\xi_2} [g]$ is antisymmetric and satisfies 
\begin{equation}
\mathcal K_{[\xi_1, \xi_2], \xi_3} + \delta_{\xi_3} \mathcal K_{\xi_1, \xi_2} + \text{cyclic}(1,2,3) =0 .
\label{cocycle condition}
\end{equation} 

Its explicit expression depends on the choice we made in the split of integrable and non-integrable parts in \eqref{infinitesimal charges}. For our prescription, choosing \eqref{Hhat} as integrable part, we have 
\begin{equation}
\begin{split}
\mathcal{K}_{\xi_1,\xi_2} [g] =& \prefwg \oint \text{d}^2\Omega \, \left[ \frac{1}{2} f_2 D_A f_1 D^A \mathring{R} + \frac{1}{2} C^{BC} f_2 D_B D_C D_D Y^D_1 \right]  \\
& + \delta_{\xi_1} (\Delta H_{\xi_2}) + \frac{1}{2}\Delta H_{[\xi_1, \xi_2]}  - (1\leftrightarrow 2)
\end{split}
\end{equation} where $\Delta H_\xi [g]$ was given in \eqref{DeltaH}. The charge algebra \eqref{eq:Algebra} closes under the modified bracket introduced in \cite{Barnich:2011mi}. Even staying on the unit round sphere $q_{AB} = \bar{q}_{AB}$ and $\delta_{\xi} q_{AB}=0$, the charge algebra differs from \cite{Barnich:2011mi} because of the shift of the charge \eqref{DeltaH} which reduces to 
\bea
\Delta H_\xi[g] = \prefwg \oint \text{d}^2 \Omega \Big[ \frac{u}{2} Y^A D_B D^B D^C C_{AC} - \frac{u}{2} Y^A D_B D_A D_C C^{BC} \Big]. 
\eea

\begin{figure}[!bth]
\vspace{-20pt}
\advance\leftskip-0.015\paperwidth % Go inwards the left margin
\subfloat[Leading BMS triangle \cite{Strominger:2017zoo}.]{
	\begin{minipage}[c][6.5cm]{0.383\paperwidth}
		\includegraphics[clip,width=0.95\textwidth]{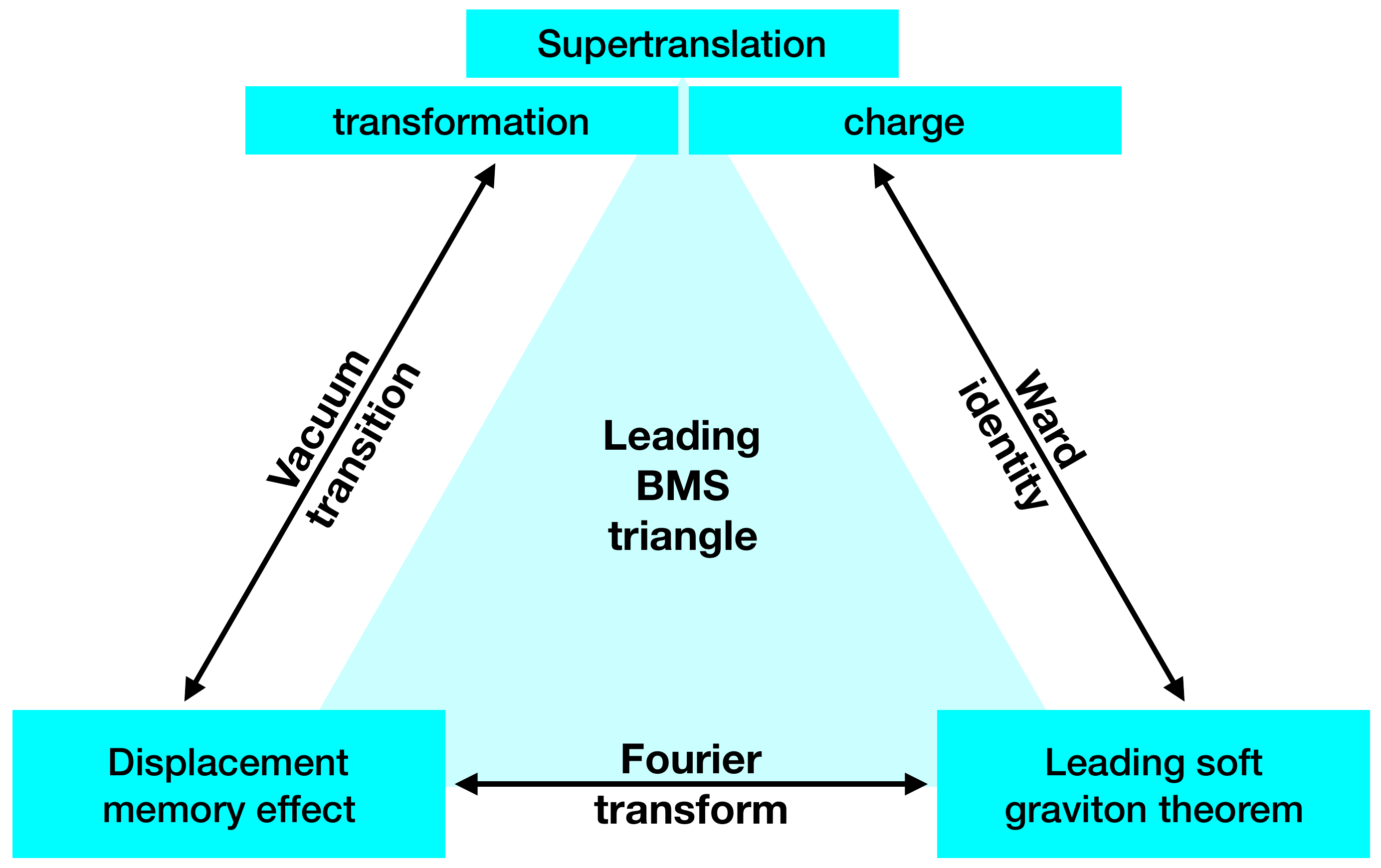}
		\label{fig:leading}
	\end{minipage}
}
\subfloat[Overleading/Subleading BMS square.]{
	\begin{minipage}[c][6.5cm]{0.383\paperwidth}
		\includegraphics[clip,width=0.99\textwidth]{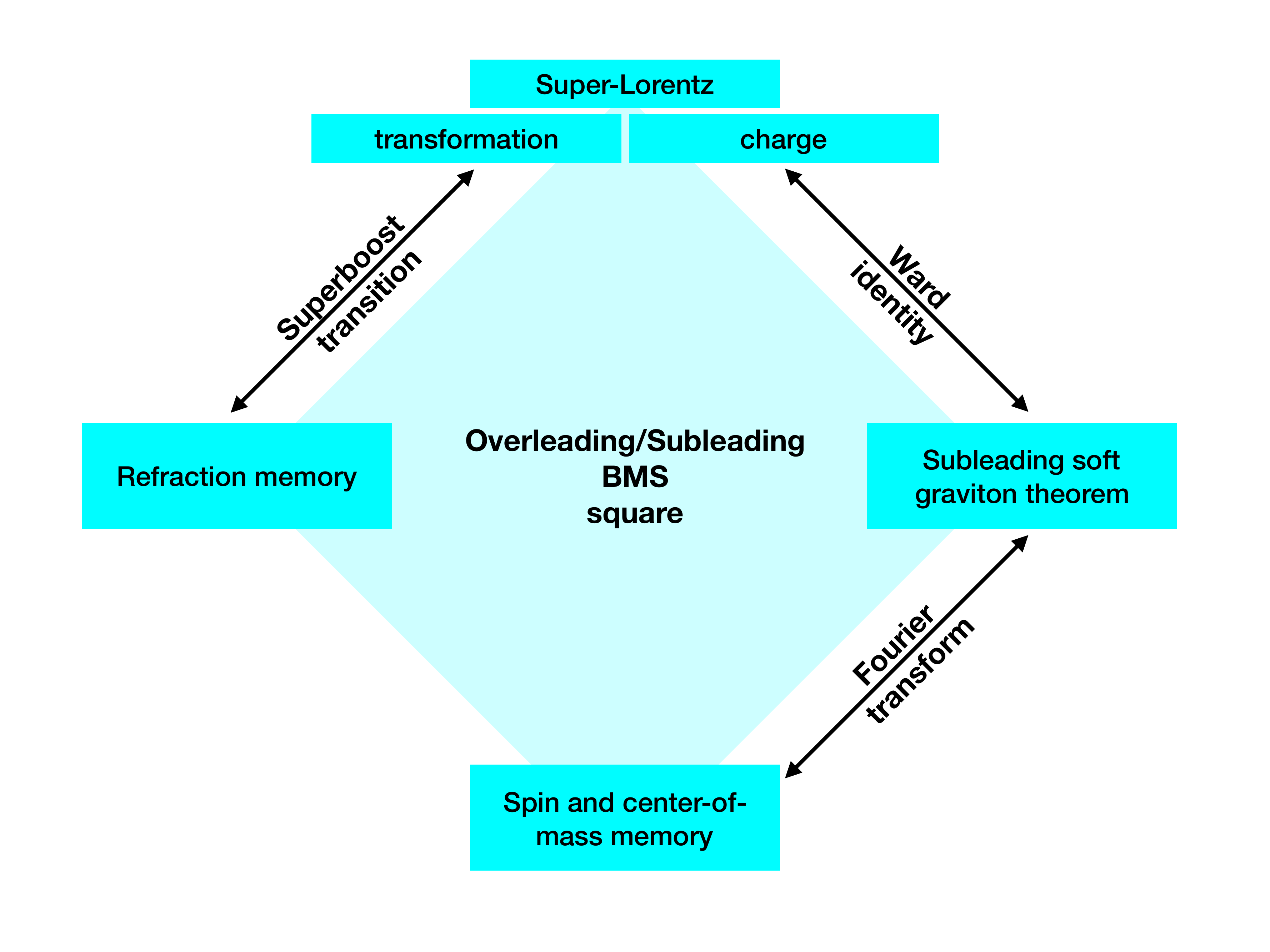}
		\label{fig:subleading}
	\end{minipage}
}
\caption{The infrared structure of asymptotically flat spacetimes at null infinity.}
\end{figure}
\section{Conclusion}

Supertranslation BMS symmetry, the leading soft graviton theorem and the displacement memory effect form three corners of a triangle describing the leading infrared structure of asymptotically flat spacetimes at null infinity \cite{Strominger:2017zoo}. The three edges of the triangle can be described in the language of vacuum transitions, Ward identities and Fourier transforms, see Figure \ref{fig:leading}. In the case of super-Lorentz BMS symmetry, one needs to distinguish the symmetry in itself which is overleading since it modifies the leading order metric at null infinity, and the charges that describe the subleading structure of gravitational fields, and in particular, their angular momentum. Two edges have been previously drawn relating super-Lorentz charges to the subleading soft graviton theorem by Ward identities \cite{Kapec:2014opa,Campiglia:2014yka,Campiglia:2015yka}, and relating the subleading soft graviton theorem to the spin effect by a Fourier transformation \cite{Pasterski:2015tva}. In this paper, we clarified how the superboost transitions lead to the refraction or velocity kick memory effect at null infinity. This suggests to describe this overleading/subleading structure by an (incomplete) square instead of a triangle as in Figure \ref{fig:subleading}. We expect that a similar overleading/subleading square structure will also appear in the description of other gauge and gravity theories.

At the technical level, we obtained a new definition of the angular momentum for standard asymptotically flat spacetimes which is consistent with the fluxes required for the subleading soft graviton theorem. We also derived the charge algebra, and described a non-linear displacement memory effect that occurs in the case of combined superboost and supertranslation transitions. The renormalized phase space that we constructed is only a first step in the definition of a general notion of asymptotic flatness. The counterterm prescription that we used requires further justification by a geometric construction. Also, superboost transitions remain to be included in the renormalized phase space. We leave these more general constructions for future endeavor.

\vspace{-10pt}
\section*{Acknowledgments}
The work of G.C. is supported by the ERC Starting Grant 335146 ``HoloBHC".  G.C. is a Research Associate and A.F. is Research Fellow of the Fonds de la Recherche Scientifique F.R.S.-FNRS (Belgium). R.R. is a FRIA (F.R.S.-FNRS) Research Fellow. We thank G. Barnich, Y. Korovin, C. Troessaert and all participants of the Solvay workshop of May 2018 on ``Infrared physics" for interesting discussions.

%\bibliography{refs}
\providecommand{\href}[2]{#2}\begingroup\raggedright \endgroup

\end{document}